\documentclass[aps,prd,twocolumn,showpacs,nofootinbib,superscriptaddress]{revtex4-1}

\usepackage{amsmath}
\usepackage{amsfonts}
\usepackage{wasysym}
\usepackage{graphicx}
\usepackage{comment}

\def\filetype{eps}

\begin{document}

\title{Extraordinary Gauge Mediation at Finite Temperature}
\author{Aaron Hanken}
\affiliation{Department of Physics and Astronomy, Rowan University, Glassboro, New Jersey 08028, USA}
\author{Ben Kain}
\author{Collin Manning}
\affiliation{Department of Physics, College of the Holy Cross, Worcester, MA 01610, USA}

\begin{abstract}
\noindent We investigate minimally completed models of extraordinary gauge mediation, which are examples of direct gauge mediation, at finite temperature both analytically and numerically.  These models may have both supersymmetry breaking and supersymmetric vacua.  Our interest is in determining the preferred vacuum.  We do so by computing the finite temperature Coleman-Weinberg potential and studying its thermal evolution.
\end{abstract} 


\pacs{12.60.Jv, 11.10.Wx}


\maketitle


\section{Introduction}

Supersymmetry is a well known candidate for physics beyond the Standard Model to be discovered at the LHC.  If true, then it must come in the form of a broken symmetry.  Intriligator, Seiberg and Shih (ISS) revitalized the idea that supersymmetry breaking could occur in a long lived, local minimum of the scalar potential, i.e.~a metastable vacuum \cite{esr, ISS1, ISrev}.  They gave simple arguments, based on R-symmetry, that we are in fact led to this possibility \cite{ISS2, ISrev}.  Nelson and Seiberg \cite{ns} have shown that in a generic model of $F$-term supersymmetry breaking R-symmetry is a necessary condition.  On the other hand, Majorana gaugino masses, necessary in realistic models, require broken R-symmetry.  This motivates us to consider models of spontaneous R-symmetry breaking \cite{dine, shih, KomShih, dinerev, CurtinShih}.

Metastable vacua and R-symmetry can be studied in renormalizable, perturbative superpotentials of the O'Raifeartaigh type \cite{ISS1, ISrev, ISS2,shih, KomShih,  ray,  ferretti, ferretti2, ako, aldmar, dienes, sun, ms, sun2, kitano0, amariti, Shih2, sun3, abdalla, zhu}.  These models must then be extended to allow for the broken supersymmetry to be communicated to the visible sector.  A standard approach for doing this is gauge mediation, which has the advantage of solving the flavor and \textit{CP} problems \cite{giudicerev}.  Cheung, Fitzpatrick and Shih made a detailed study of gauge mediation for a large family of models using the most generic R-symmetric superpotential that follows from $F$-term supersymmetry breaking and referred to them as models of extraordinary gauge mediation (EOGM):
\begin{equation} \label{eogmsp0}
	W_{\text{EOGM}} = m_{ij} \phi_i \widetilde{\phi}_j + \lambda_{ij} x\phi_i \widetilde{\phi}_j,
\end{equation}
where the $\phi_i$ and $\widetilde{\phi}_i$ are messenger fields in the $5\oplus \bar{5}$ representation of $SU(5)$ and $x$ is a singlet spurion field which obtains scalar and $F$-term VEVs and breaks supersymmetry.  By combing models of supersymmetry breaking and EOGM one can construct models of direct gauge mediation, i.e.~models in which the messengers help both break supersymmetry and mediate it.  The minimal completions of EOGM \cite{eogm} do precisely this and are obtained by simply adding a term proportional to $x$ to (\ref{eogmsp0}):
\begin{equation} \label{eogmsp}
	W = fx +  m_{ij} \phi_i \widetilde{\phi}_j + \lambda_{ij} x\phi_i \widetilde{\phi}_j.
\end{equation}
It is these models that we study in this paper.

We consider R-symmetric minimal completions of EOGM with superpotentials of the form (\ref{eogmsp}) and canonical K\"ahler potentials.  Since the R-symmetry requires $x$ to have R-charge $R(x) = 2$, vacua with $\langle x \rangle \neq 0$ spontaneously break R-symmetry.  The superpotential (\ref{eogmsp}), at treelevel, spontaneously breaks supersymmetry along the flat direction \cite{eogm}
\begin{equation} \label{flat}
	\phi_i = \widetilde{\phi}_i = 0, \qquad x = \text{arbitrary}.
\end{equation}
The location of the vacuum,  $\langle x\rangle$, can be determined by including one-loop Coleman-Weinberg corrections.  One can only compute such corrections at points where the scalar potential is stable (i.e.~points where the eigenvalues of the scalar mass squared matrix are all non-negative).  In the $\phi_i = \widetilde{\phi}_i = 0$ direction, the scalar potential is stable in some range
\begin{equation}
	x_{\text{min}} < |x| < x_{\text{max}},
\end{equation}
where $x_{\text{min}}$ and $x_{\text{max}}$ are model dependent and could be zero and infinity, respectively.  In \cite{eogm}, models of EOGM, including their minimal completions, were classified into three types (types I, II and III).  In the following sections we adhere to these classifications, give their definitions and elaborate on their details.

The models we study can have a supersymmetry and R-symmetry breaking vacuum in the $\phi_i = \widetilde{\phi}_i = 0$ direction at some $\langle x \rangle \neq 0$, which we shall refer to as the EOGM vacuum.  However, this may not be the only vacuum in the theory.  It is possible that supersymmetric vacua, including at the end of a runaway direction, exist \cite{eogm}.  An important question to ask, then, is which vacuum is favored, as the viability of a model depends in part on the supersymmetry breaking vacuum being the favored vacuum.  This question may be answered by studying the thermal evolution of the scalar potential.  Including one-loop finite temperature corrections, the resulting scalar potential is assumed to be the scalar potential relevant for the expanding Universe.  At very high temperatures, as was the case during the early hot Universe, there is usually a single, global minimum at the origin of field space.  As the temperature is lowered, which we assume to mimic the expanding and cooling of the Universe, new vacua emerge.  By studying this thermal evolution, we can determine the favored zero temperature vacuum.

The thermal evolution of supersymmetry breaking has been investigated previously for the ISS model \cite{abel, abel2, craig, fischler, anguelova} and for O'Raifeartaigh-like models \cite{katz, moreno, arai, dalianis}.  We study the thermal evolution of minimal completions of EOGM.  Such a study was first made by Katz \cite{katz} analytically.  Specific models, which are closely related to type I models, were analyzed numerically by Moreno and Schaposnik \cite{moreno} and Arai, Kobyashi and Sasaki \cite{arai}.  In the next three sections, we study the thermal evolution of type I, type II and type III models.  In each section, after general remarks, we consider a specific example in detail using both numerical and analytical techniques.  In this way we see our work as complementary to \cite{katz}.  For the $\phi_i = \widetilde{\phi}_i = 0$ direction we analyze the thermal evolution of the scalar potential and determine the parameter space in which an R-symmetry breaking minimum exists.  When appropriate, we then give an analytical study at zero temperature using the effective K\"ahler potential \cite{ISS1, effK, effK2}.  Finally, we consider the existence of supersymmetric vacua, such as at the end of a runaway direction, and the important question of which is the favored vacuum.  We conclude in section \ref{conclusion}.  In the Appendix we give the complete mass matrices for the models under study.  We assume throughout that the Standard Model couplings are much weaker than the messenger couplings in the EOGM model, allowing us to ignore Standard Model interactions \cite{eogm, katz}, and that the Universe cools adiabatically \cite{fischler, katz}.


\section{Type I}
\label{sec:typeI}

Type I models are the most straightforward generalization of O'Raifeartaigh models.  They are defined by $\det m \neq 0$, from which follows $\det \lambda = 0$.  The pseudomoduli space (\ref{flat}) is stable in the range
\begin{equation} \label{xminmaxI}
	|x| < x_{\text{max}},
\end{equation}
for some model dependent $x_{\text{max}}$.  They have vanishing leading order gaugino masses and, in this sense, are somewhat pathological \cite{eogm}.  At very high temperatures, the global minimum of the potential is at the origin of field space.  If the Universe begins in this global minimum, ending up in the EOGM vacuum (\ref{flat}) can be cosmologically favored.  At high temperatures, type I models can possess an additional, local vacuum separated from the origin.  If the Universe instead begins in this vacuum, it can be the supersymmetric vacuum at the end of a runaway direction that is cosmologically favored \cite{katz, moreno}.

The specific type I model we analyze is Shih's model of spontaneous R-symmetry breaking \cite{shih}.  This model was originally presented as the simplest R-symmetric O'Raifeartaigh-type model exhibiting spontaneous R-symmetry breaking and not as a model of gauge mediation.  Here we incorporate Shih's model into the minimally completed EOGM structure by promoting the singlet fields to messengers transforming in the $5 \oplus \bar{5}$ representation of $SU(5)$.  This is easily done using an equivalent form for the $m_{ij}$ and $\lambda_{ij}$ matrices used in \cite{shih}:
\begin{equation} \label{typeImat}
	m_{ij} = \left(
		\begin{array}{ccc}
			m_1 & 0 & 0 \\
			0 & m_2 & 0 \\
			0 & 0 & m_1
		\end{array} \right),
	\qquad
	\lambda_{ij} = \left(
		\begin{array}{ccc}
			0 & \lambda & 0 \\
			0 & 0 & \lambda \\
			0 & 0 & 0
		\end{array} \right),
\end{equation}
which are seen to satisfy $\det m \neq 0$ and $\det \lambda = 0$.  Placing these matrices into (\ref{eogmsp}) we obtain the superpotential
\begin{equation} \label{typeIsp}
	W = fx + m_1 (\phi_1 \widetilde{\phi}_1 + \phi_3\widetilde{\phi}_3) + m_2 \phi_2 \widetilde{\phi}_2 + \lambda x (\phi_1 \widetilde{\phi}_2 + \phi_2 \widetilde{\phi}_3).
\end{equation}
From this superpotential we can see that this model is R-symmetric with R-charge assignments
\begin{equation}
\begin{gathered}
	R(x) = 2, \quad
	R(\phi_1) = - R(\widetilde{\phi}_2) = \alpha, \\
	R(\phi_2) = - R(\widetilde{\phi}_3) = 2 + \alpha, \quad
	R(\phi_3) = 4 + \alpha, \\
	R(\widetilde{\phi}_1) = 2 - \alpha,
\end{gathered}
\end{equation}
for arbitrary $\alpha$.  The treelevel scalar potential is
\begin{equation} \label{treeI}
\begin{split}
	V_0 &= |f + \lambda (\phi_1 \widetilde{\phi}_2 + \phi_2 \widetilde{\phi}_3)|^2 + |m_1 \widetilde{\phi}_1 + \lambda x \widetilde{\phi}_2|^2 + |m_1 \phi_1| ^2 \\
	&\qquad+ |m_2 \widetilde{\phi}_2 + \lambda x \widetilde{\phi}_3|^2 + |m_2 \phi_2 + \lambda x \phi_1|^2 + |m_1 \widetilde{\phi}_3|^2 \\
	&\qquad+ |m_1 \phi_3 + \lambda x \phi_2|^2.
\end{split}
\end{equation}
We assume all couplings to be real and positive without loss of generality, which can always be obtained by rotating the phases of the fields.  

The scalar potential (\ref{treeI}) has supersymmetry breaking extrema at (\ref{flat}).  Since we are interested in the vacuum structure it is important to determine the stability conditions such that the extrema (\ref{flat}) is a minimum.  These stability conditions also inform us of the parameter space in which the one-loop Coleman-Weinberg corrections can be reliably computed.  As we shall see, the Coleman-Weinberg corrections lift the flat direction (\ref{flat}), yielding a local, metastable minimum at some $\langle x\rangle $, so that we have a degenerate pseudomoduli space of supersymmetry breaking vacua in the $\phi_i = \widetilde{\phi}_i = 0$ direction, parameterized by $x$, the pseudomodulus.  If $\langle x \rangle \neq 0$ R-symmetry is spontaneously broken since $x$ is charged under the R-symmetry.  This minima is local and metastable because the global, supersymmetry preserving minimum exists at the end of a runaway direction, whose study we postpone until the end of this section.

The $\phi_i = \widetilde{\phi}_i = 0$ direction is stable if the scalar mass squared matrix eigenvalues in this direction are all non-negative.  The general mass matrices for this model are given in section \ref{mBmFI} in the Appendix.   The stability conditions are conveniently parameterized in terms of the dimensionless parameters \cite{shih}
\begin{equation} \label{rydefI}
	r \equiv \frac{m_2}{m_1}, \qquad
	y \equiv \frac{\lambda f}{m_1 m_2}.
\end{equation}
One finds that the $\phi_i = \widetilde{\phi}_i = 0$ direction is stable when \cite{shih}
\begin{equation} \label{stablepspaceI}
	y < 1, \qquad
	|x| < x_{\text{max}} = \frac{m_1}{\lambda} \frac{1-y^2}{2 y}.
\end{equation}
As expected, these stability conditions are consistent with (\ref{xminmaxI}).  As mentioned above, they also give the parameter space in which the one-loop Coleman-Weinberg corrections can be reliably computed.

Loop corrections play an important role in analyzing the vacuum structure of the model.  As previously mentioned, there exists a degenerate pseudomoduli space of supersymmetry breaking vacua, parameterized by $x$, at $\phi_i = \widetilde{\phi}_i = 0$.  If loop corrections are included this degeneracy is lifted.  The one-loop Coleman-Weinberg correction to the scalar potential has both zero temperature and finite temperature contributions:
\begin{equation}
	V_1 = V_1^{(0)} + V_1^{(T)}.
\end{equation}
This may be added to the tree level scalar potential (\ref{treeI}) to obtain the complete one-loop finite temperature scalar potential, $V = V_0 + V_1$, where, in the $\phi_i = \widetilde{\phi}_i = 0$ direction, $V_0 = f^2$ is constant.  The Coleman-Weinberg formulas are well known:
\begin{subequations} \label{CWeqs}
\begin{align}
	V^{(0)}_1 &= \frac{1}{64\pi^2} \sum_i \left[ (m^2_{S,i})^2 \ln \frac{m^2_{S,i}}{\Lambda^2} - (m^2_{F,i})^2 \ln \frac{m^2_{F,i}}{\Lambda^2} \right] \label{CWeqsa}\\
	V_1^{(T)} &=  
	\quad\frac{T^4}{2\pi^2} \sum_i \notag\\ &\times\int\limits_0^\infty dq \, q^2 
	\biggl\{
	 \ln \left[ 1 - \exp \left( - \sqrt{q^2 + m_{S,i}^ 2 / T^2} \right) \right] \notag\\
	&\qquad\qquad - \ln \left[ 1 + \exp \left( - \sqrt{q^2 + m_{F,i}^2 / T^2} \right) \right] \biggr\}, \label{CWeqsb}
\end{align}
\end{subequations}
where $m_{S,i}^2$ and $m_{F,i}^2$ are the $i$th eigenvalues of the scalar and fermion mass squared matrices, $T$ is the temperature and $\Lambda$ is a momentum cutoff.  Since the eigenvalues cannot be determined analytically in this model we have two options: An exact numerical analysis or introducing approximations such than an analytical analysis is possible.  We pursue both options.

Both analyses are simplified if we move to dimensionless quantities.  We define
\begin{equation} \label{dimeqsI}
\begin{gathered}
	\overline{V} \equiv \frac{1}{m_1^4} V, \qquad
	\overline{x} \equiv \frac{\lambda}{m_1} x, \qquad
	\overline{\phi}_i \equiv \frac{\lambda}{m_1} \phi_i, \\
	\overline{\widetilde{\phi}}_i \equiv \frac{\lambda}{m_1} \widetilde{\phi}_i, \qquad
	\overline{T} \equiv \frac{1}{m_1} T,  \qquad
	\overline{\Lambda} \equiv \frac{1}{m_1} \Lambda, 	\\
	\overline{m}_S^2 \equiv \frac{1}{m_1^2} m_S^2, \qquad
	\overline{m}_F^2 \equiv \frac{1}{m_1^2} m_F^2,
\end{gathered}
\end{equation}
which accompany $r$ and $y$ defined in (\ref{rydefI}).  In terms of these quantities the treelevel scalar potential (\ref{treeI}) simplifies to
\begin{equation} \label{treeI2}
\begin{split}
	\lambda^2\overline{V}_0 &= |yr + \overline{\phi}_1 \overline{\widetilde{\phi}}_2 + \overline{\phi}_2 \overline{\widetilde{\phi}}_3|^2 + | \overline{\widetilde{\phi}}_1 + \overline{x} \overline{\widetilde{\phi}}_2|^2 + | \overline{\phi}_1| ^2 \\
	&\qquad+ |r\overline{\widetilde{\phi}}_2 + \overline{x} \overline{\widetilde{\phi}}_3|^2 + |r\overline{\phi}_2 + \overline{x} \overline{\phi}_1|^2  \\
	&\qquad+ |\overline{\widetilde{\phi}}_3|^2 + |\overline{\phi}_3 + \overline{x} \overline{\phi}_2|^2,
\end{split}
\end{equation}
and in the $\phi_i = \widetilde{\phi}_i = 0$ direction becomes $\overline{V}_0 = (yr/\lambda)^2 $.  The stability conditions (\ref{stablepspaceI}) become $y<1$ and $\overline{x} < (1-y^2)/2y$ and the dimensionless mass matrices depend on $r$, $y$ and the dimensionless fields, but not $\lambda$.  The Coleman-Weinberg formulas (\ref{CWeqs}) become
\begin{subequations} \label{CWeqs2}
\begin{align}
	\overline{V}^{(0)}_1 &= \frac{1}{64\pi^2} \sum_i \left[ (\overline{m}^2_{S,i})^2 \ln \frac{\overline{m}^2_{S,i}}{\overline{\Lambda}^2} - (\overline{m}^2_{F,i})^2 \ln \frac{\overline{m}^2_{F,i}}{\overline{\Lambda}^2} \right] \label{CWeqsa2}\\
	\overline{V}_1^{(T)} &=  \frac{\overline{T}^4}{2\pi^2} \sum_i \notag\\
	&\times \int\limits_0^\infty dq \, q^2 \biggl\{
	 \ln \left[ 1 - \exp \left( - \sqrt{q^2 + \overline{m}_{S,i}^ 2 / \overline{T}^2} \right) \right]\notag\\
	 &\qquad\qquad
	- \ln \left[ 1 + \exp \left( - \sqrt{q^2 + \overline{m}_{F,i}^2 / \overline{T}^2} \right) \right] \biggr\}. \label{CWeqsb2}
\end{align}
\end{subequations}

\subsection{Numerical}

A numerical analysis of Shih's model of spontaneous R-symmetry breaking \cite{shih} at finite temperature was undertaken by Moreno and Schaposnik \cite{moreno}.  While Moreno and Schaposnik did not incorporate Shih's model into minimal completions of EOGM, as we are doing here, our numerical results mostly reproduce theirs.  Still, we include these results for completeness and also as a means to give our first and most comprehensive description of the numerical methods we employ, methods we use again for the type II and III models studied below.

Since our interest here is in minima of the scalar potential in the $\phi_i = \widetilde{\phi}_i = 0$ direction, we do not need to include the treelevel contribution, $\overline{V}_0 = (yr/\lambda)^2 $, which in this direction is constant and has no effect.  However, we include it for future comparison with the runaway direction studied in section \ref{runawayI}.  We must specify, then, a value for $\lambda$, which we take to be $\lambda = 1$, but note that the position of minima does not depend on this choice.   The location of the minima in the $\phi_i = \widetilde{\phi}_i = 0$ direction depends on $\bar{x}$, $r$, $y$, $\overline{T}$ and the cutoff $\overline{\Lambda}$.

\begin{figure*}
		\includegraphics[width=6in]{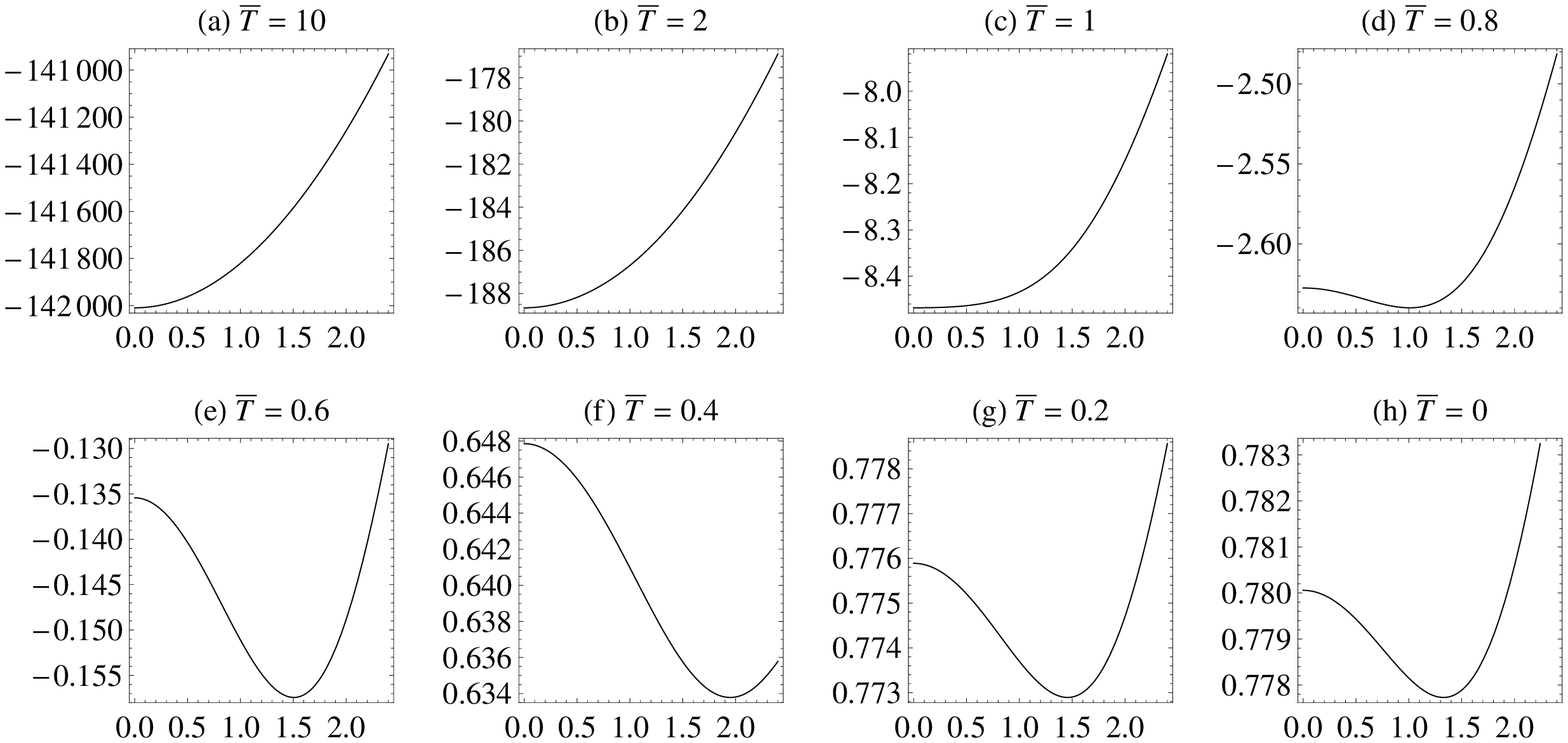}[h]
	\caption{$\overline{V}$ as a function of $|\bar{x}|$ is plotted for $r=4$, $y=0.2$, $\lambda = 1$,  $\overline{\Lambda}=1$ and various temperatures for the type I model.  The treelevel potential is stable and the one loop corrections are computable for $ |\overline{x}| < \overline{x}_{\text{max}} = 2.4$.  The zero temperature minimum in plot (h) is located at $\langle |\bar{x}| \rangle_0 = 1.327$.  The second order phase transition occurs at critical temperature $\overline{T}_c = 0.96$.  Neither of these quantities depend on $\lambda$. }
\label{fig:V1I}
\end{figure*}

We may now use the Coleman-Weinberg formulas (\ref{CWeqs2}) and numerically compute $\overline{V}_1$.  For a typical set of parameters ($r=4$, $y=0.2$, $\lambda = 1$ and $\overline{\Lambda}=1$ \cite{shih}) we have plotted $\overline{V}$ as a function of $|\bar{x}|$ for various temperatures in figure \ref{fig:V1I}, which reproduces plots in \cite{shih,moreno}.
  In figure \ref{fig:V1I}(a) the expected high temperature minimum at the origin of field space is apparent. Figures \ref{fig:V1I}(b--g) show the thermal evolution of the potential and in figure \ref{fig:V1I}(h) is plotted the zero temperature scalar potential, clearly showing an R-symmetry breaking minimum, which is the EOGM vacuum and is located at \cite{shih}
\begin{equation} \label{numxminI}
	\langle |\bar{x}|\rangle_0 = 1.327
\end{equation}
(the subscript 0 indicates that this is the zero temperature minimum).  As the temperature is lowered the potential undergoes a second order phase transition, the minimum relocates away from the origin and R-symmetry is broken.  The critical temperature for the phase transition is
\begin{equation}
	\overline{T}_c = 0.96.
\end{equation}

\begin{figure*}[t]
		\includegraphics[width=6in]{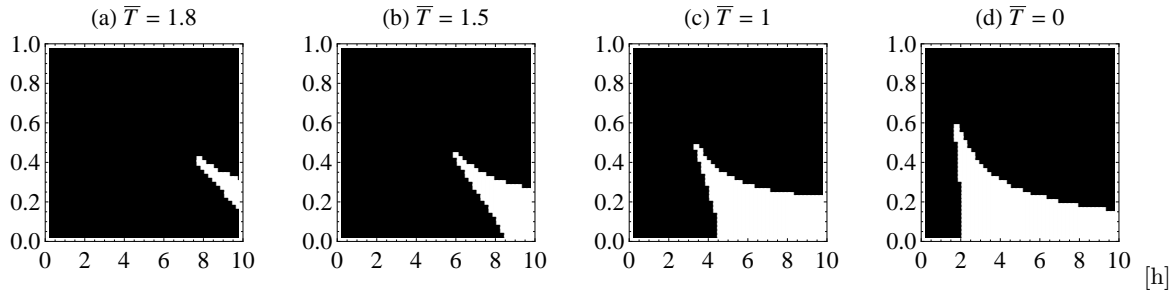}[h]
	\caption{$y$ as function of $r$ is plotted for cutoff $\overline{\Lambda} =1$ and various temperatures for the type I model.  The white region is the computable parameter space in which an R-symmetry breaking minimum exists.}
\label{fig:pspaceI}
\end{figure*}

The minima in figure \ref{fig:V1I} are for a single set of parameters $r$, $y$, $\overline{T}$ and $\overline{\Lambda}$.  We can scan through these and plot the computable parameter space in which an R-symmetry breaking minimum exists.  The results are shown in figure \ref{fig:pspaceI}, which reproduces plots in \cite{shih,moreno},
 where each is plotting $y$ as a function of $r$ for a given temperature.  The white region is where an R-symmetry breaking minimum exists.  We stress that figure \ref{fig:pspaceI} is the computable parameter space because we restricted our search to the region (\ref{stablepspaceI}) where the Coleman-Weinberg corrections, at zero and finite temperature, can be reliably computed.


\subsection{Analytical}
\label{typeIanalytical}

The zero temperature Coleman-Weinberg correction (\ref{CWeqsa}) depends on the eigenvalues of the mass matrices, which in this model cannot be computed analytically.  In the previous subsection we computed them numerically.  In this subsection we employ approximations and obtain approximate analytical results.

In the small supersymmetry breaking limit, the one-loop scalar potential at zero temperature may be obtained through the effective K\"ahler potential \cite{ISS1}:
\begin{equation}
	K_{\text{eff}} = K_0 + K_1.
\end{equation}
$K_0$ is the treelevel K\"ahler potential, which we take to be canonical (something we had implicitly been assuming above), and \cite{effK, effK2, ISS1}
\begin{equation} \label{K1eqI}
\begin{split}
	K_1 &= 
	- \frac{1}{32\pi^2} \text{Tr} \left( W^\dagger_{ik}W^{kj} \ln \frac{W^\dagger_{ik}W^{kj}}{\Lambda^2} \right)\\
	&=
	- \frac{1}{64\pi^2} \text{Tr} \left( m_F^2 \ln \frac{m_F^2}{\Lambda^2} \right)
\end{split}
\end{equation}
is the one-loop correction.  In the $\phi_i = \widetilde{\phi}_i = 0$ direction, the one-loop scalar potential at zero temperature in the small supersymmetry breaking limit may be obtained from
\begin{equation} \label{VKeffI}
	V = \left( \frac{\partial^2 K_{\text{eff}}}{\partial x \partial x^*} \right)^{-1} \left| \frac{\partial W}{\partial x} \right|^2,
\end{equation}
as an alternative to using the Coleman-Weinberg formula (\ref{CWeqsa}).

The formula (\ref{K1eqI}) requires computing eigenvalues of the mass matrices.  To avoid this we expand $K_1$.  Since the superpotential is R-symmetric, the one-loop correction (\ref{K1eqI}) will contain only R-symmetric terms and be of the form
\begin{equation} \label{K1expansionI}
	K_1 = k_0 + k_2|x|^2 + k_4|x|^4 + k_6|x|^6 + \cdots.
\end{equation}
The coefficients $k_i$ can be computed analytically without knowledge of the mass matrix eigenvalues \cite{lalak, bfk}.  We assume that we can safely truncate the expansion (\ref{K1expansionI}) after the $|x|^6$ term.  Using this truncated expansion in (\ref{VKeffI}), the one-loop corrected zero temperature scalar potential is then approximated by \cite{lalak, bfk}
\begin{equation} \label{rsymspI}
	V^{(0)} \approx |f|^2 -4|f|^2k_{4}|x|^2 -9|f|^2k_{6}|x|^4.
\end{equation}
The scalar potential (\ref{rsymspI}) will have an R-symmetry breaking minimum ($\langle |x| \rangle_0 \neq 0$) if
\begin{equation} \label{kcriteriaI}
	k_4 > 0, \qquad k_6 < 0,
\end{equation}
and it will be located at 
\begin{equation} \label{keffminI}
	\langle |x|^2 \rangle_0 =  \frac{2 k_4}{9|k_6|}.
\end{equation}

As before, it proves convenient to move to dimensionless quantities:
\begin{equation}
	\bar{k}_4 \equiv \frac{m_1^2}{\lambda^4} k_4, \qquad
	\bar{k}_6 \equiv  \frac{m_1^4}{\lambda^6} k_6.	
\end{equation}
The one-loop corrected zero temperature scalar potential (\ref{rsymspI}) becomes
\begin{equation} \label{rsymsp2}
	\overline{V}^{(0)} \approx |yr/\lambda|^2 -4|yr|^2 \bar{k}_{4}|\overline{x}|^2 -9|yr|^2\bar{k}_{6}|\overline{x}|^4.
\end{equation}
and the R-symmetry breaking minimum (\ref{keffminI}) is located at
\begin{equation} \label{keffminI2}
	\langle |\overline{x}|^2 \rangle_0 =  \frac{2 \bar{k}_4}{9|\bar{k}_6|}.
\end{equation}

$\bar{k}_4$ and $\bar{k}_6$ may be computed using, for example, the methods developed in \cite{bfk}.  We find
 \cite{shih, lalak, bfk}
\begin{subequations} \label{typeIk4k6}
\begin{align}
	\bar{k}_4 &= \frac{5}{16 \pi^2 } \frac{1 + 2r^2 -3r^4 + r^2(3+r^2)\ln r^2}{(r^2 - 1)^3} \\
\begin{split}
	\bar{k}_6 &=  - \frac{5}{16 \pi^2} \\
	&\times\frac{1 + 27r^2 - 9r^4 - 19r^6 + 6r^2(2+5r^2 + r^4)\ln r^2}{3(r^2-1)^5}.
\end{split}
\end{align}
\end{subequations}
These formulas obey (\ref{kcriteriaI}), and consequently R-symmetry is spontaneously broken, for \cite{shih}
\begin{equation}
	r > 2.12,
\end{equation}
in agreement with figure \ref{fig:pspaceI}(d).  The location of the zero temperature vacuum is given by (\ref{keffminI2}), which depends only on $r$.  For $r=4$ we find $\langle | \bar{x} | \rangle_0 = 1.255$, which agrees with the numerical result (\ref{numxminI}) to two digits.

In this subsection we used analytical approximations for the zero temperature Coleman-Weinberg formula (\ref{CWeqsa}) that did not require computing the eigenvalues of the mass matrices.  There also exists analytical approximations for the finite temperature Coleman-Weinberg formula (\ref{CWeqsb}), but only for high temperatures ($\overline{T} \gg 1$).  A look at figure \ref{fig:V1I} shows us that this is beyond the temperatures of interest.  Consequently, we will not study this model at finite temperature analytically.


\subsection{Runaway Direction}
\label{runawayI}

So far we have been studying supersymmetry and R-symmetry breaking vacua in the $\phi_i = \widetilde{\phi}_i = 0$ direction. It is possible that a supersymmetric vacuum exists when one or more of the fields is sent to infinity, i.e.~there may exist a runaway direction.  If a runaway direction exists, one must check which vacuum is favored as the validity of a model relies on the likelihood of ending up in the supersymmetry breaking vacuum.

To the best of our knowledge, no one has presented a complete study of the possibility of escaping to the runaway direction.  This is not unreasonable as it is very difficult to investigate the complete scalar potential when all fields are nonzero.  Instead, the standard analysis \cite{abel, moreno, arai} is to choose a single path (of possibly many or infinite paths) to the runaway direction and to study the likelihood of the system escaping to the runaway via this single path.

At very high temperatures, the origin of field space is the global minimum of the potential.  In the type I model we are studying, as the temperature drops, local and global vacua develop away from the origin.  We take the initial temperature of the system to be equal to the reheating temperature and we ask which vacuum---the supersymmetry breaking EOGM vacuum or the supersymmetric vacuum at the end of the runaway direction---is favored at zero temperature.  Further, we make the standard assumption that the finite temperature corrections give an approximate modeling of the thermal evolution of the Universe.  As the Universe expands and cools to the present era, its temperature drops to roughly zero and we are investigating which vacuum we expect to end up in.

The scalar potential (\ref{treeI2}) goes to zero in the limit \cite{shih}
\begin{equation} \label{rundirI}
\begin{gathered}
	\overline{x} \rightarrow \left( \frac{2}{y} \overline{\phi}_3^2 \right)^{1/3}, \qquad
	\overline{\phi}_1 \rightarrow \left(\frac{y^2 r^3}{4} \frac{1}{\overline{\phi}_3} \right)^{1/3}, \\
	\overline{\phi}_2 \rightarrow - \left(\frac{y}{2} \overline{\phi}_3 \right)^{1/3},  \qquad
	\overline{\phi}_3 \rightarrow \infty, \\
	\overline{\widetilde{\phi}}_1 \rightarrow \phi_3, \qquad
	\overline{\widetilde{\phi}}_2 \rightarrow \phi_2, \qquad
	\overline{\widetilde{\phi}}_3 \rightarrow \phi_1,
\end{gathered}
\end{equation}
and thus (\ref{rundirI}) is a runaway direction.  We wish to find a path that begins at $\langle \overline{x}\rangle_T$, the R-symmetry breaking vacuum at temperature $T$ in the $\phi_i = \widetilde{\phi}_i = 0$ direction, and ends at (\ref{rundirI}).  Further, it is important that this path be stable, i.e.~the scalar mass squared matrix along this path has no non-negative eigenvalues.  Unfortunately, most paths have some portion that is unstable \cite{arai}.  The path obtained by changing the arrows in (\ref{rundirI}) to equal signs lies on the surface defined by the scalar and fermion mass squared matrices being equal, $m_S^2 = m_F^2$.  Since the fermion mass matrix is hermitian, and thus its square always has non-negative eigenvalues, this assures us that this path is always stable \cite{arai}.  Unfortunately, the starting point, $\langle \overline{x}\rangle_T$, which lies along $\phi_i = \widetilde{\phi}_i = 0$, is not on this path.

There are different methods for connecting $\langle \overline{x}\rangle_T$ to the runaway \cite{moreno,arai}.  We use the method developed in \cite{moreno}: The path is given by
\begin{equation} \label{runpathI}
\begin{gathered}
	\overline{x} = \left( \frac{2}{y} \overline{\phi}_3^2 \right)^{1/3} + [1-h(\overline{\phi}_3)]\langle \overline{x} \rangle _T , \\
	\overline{\phi}_1  = h(\overline{\phi}_3) \left(\frac{y^2 r^3}{4} \frac{1}{\overline{\phi}_3} \right)^{1/3}, \qquad
	\overline{\phi}_2 = - \left(\frac{y}{2} \overline{\phi}_3 \right)^{1/3} \\
	\overline{\widetilde{\phi}}_1 = \overline{\phi}_3, \qquad
	\overline{\widetilde{\phi}}_2 = \overline{\phi}_2, \qquad
	\overline{\widetilde{\phi}}_3 = \overline{\phi}_1,
\end{gathered}
\end{equation}
where $h = h(\overline{\phi}_3)$ is a function that satisfies
\begin{equation} \label{hreqI}
	h(0) = 0, \quad
	\lim_{\overline{\phi}_3 \rightarrow 0}
	\overline{\phi}_3^{\,-1/3} h(\overline{\phi}_3) = 0, \quad
	\lim_{\overline{\phi}_3 \rightarrow \infty} h(\overline{\phi}_3) = 1.
\end{equation}
For such a function the path (\ref{runpathI}) begins at $\langle \overline{x} \rangle_T$ and ends at the runaway for $ 0 < \phi_3 < \infty$.  There are many possible choices for $h(\phi_3)$.  All choices we have tried leave some part of the path unstable \cite{arai}.  We choose
\begin{equation} \label{hI}
	h(\overline{\phi}_3) = \tanh(c \overline{\phi}_3),
\end{equation}
for some constant $c$.  Regardless of the value of $c$, we find that the unstable region is always a small segment of the path between $\phi_3 = 0$ and the first peak of the scalar potential along the path.  We somewhat arbitrarily set $c=10$ (arbitrarily choosing $c$ is justified because, again, we use the approach that we construct a single path to the runaway and study the likelihood of escaping to the runaway via this one path).

\begin{figure*}
	\includegraphics[width=6in]{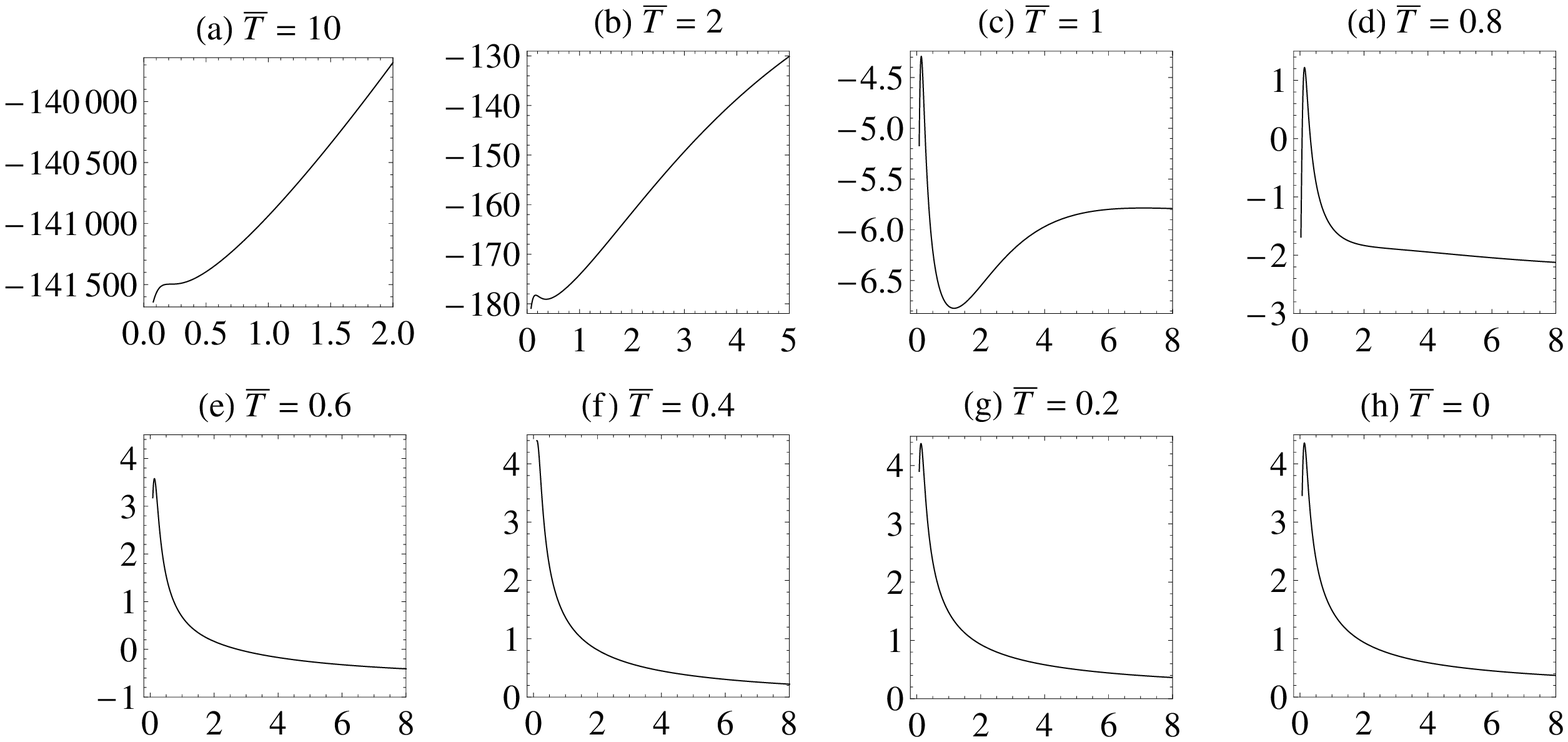}[h]
	\caption{The scalar potential, $\overline{V}$, along the path (\ref{runpathI}) is plotted as a function of $\overline{\phi}_3$ for $r=4$, $y=0.2$, $\lambda = 1$, cutoff $\overline{\Lambda}=1$ and various temperatures for the type I model.}
\label{fig:runawayI}
\end{figure*}

Having determined a path, we can compute the loop corrected scalar potential along this path.  To do this, we substitute the path (\ref{runpathI}) into the scalar and fermion mass squared matrices given in the Appendix and then use the Coleman-Weinberg formulas.  Using the same parameters ($r=4$, $y=0.2$, $\lambda = 1$ and $\overline{\Lambda}=1$) and temperatures used in figure \ref{fig:V1I}, the result is shown in figure \ref{fig:runawayI}.
 In figure \ref{fig:runawayI} we have only plotted that part of the scalar potential that is stable and thus computable.

The vertical axes of the plots in figure \ref{fig:runawayI} represent $\overline{\phi}_i = \overline{\widetilde{\phi}}_i = 0$ and $\overline{x} = \langle |\overline{x}|\rangle_T$.  This point is stable, since it lies along (\ref{xminmaxI}), and is the same point represented by the vertical axes of the plots in figure \ref{fig:V1I}.  As mentioned, parts of the chosen path connecting this point and the first peak in the scalar potential can be unstable and thus are missing in some of the plots in figure \ref{fig:runawayI}.  Still, we may be able to infer what is happening.  By comparing figure \ref{fig:runawayI} with figure \ref{fig:V1I} we see that the point $\overline{\phi}_i = \overline{\widetilde{\phi}}_i = 0$ and $\overline{x} = \langle |\overline{x}|\rangle_T$, which has not been plotted in figure \ref{fig:runawayI}, always has lower potential energy than the nearby part of the path to the runaway.  Further, we can reliably determine a local maximum near this point in all but figure \ref{fig:runawayI}(f).  Since the scalar potential is continuous, it must rise from $\langle |\overline{x}|\rangle_T$ to connect to that part plotted in figure \ref{fig:runawayI}.  (Since this is true for figure \ref{fig:runawayI}(f), it implies that figure \ref{fig:runawayI}(f) also has a local maximum.)  It appears, then, that there is always a potential barrier between $\langle | \overline{x} |\rangle_T$ and the runaway.  If we were to start in the $\langle |\overline{x}|\rangle_T$ vacuum, this potential barrier blocks the system from escaping to the supersymmetric vacuum \cite{abel, abel2, craig, fischler, moreno, arai}.  As predicted by Katz \cite{katz, moreno}, we find also that there exists a high temperature minimum, visible in figures \ref{fig:runawayI}(b) and (c), that disappears by zero temperature.  If we were to find ourselves in this minimum, instead of the one at the origin, then the supersymmetric vacuum at the end of the runaway is the preferred zero temperature vacuum.  However, by comparing figures \ref{fig:V1I} and \ref{fig:runawayI}, we find that $\langle |\overline{x} |\rangle_T$ in figure \ref{fig:V1I} has lower potential energy than this high temperature minimum and thus is favored \cite{arai}.  As such, we conclude that the supersymmetry breaking EOGM vacuum $\langle | \overline{x}| \rangle_0$ is the preferred zero temperature vacuum.

Since the EOGM vacuum is only metastable, one should also consider the possibility of a first order phase transition, i.e.~tunneling through the potential barrier to the supersymmetric vacuum.  This has been presented in detail elsewhere \cite{shih, katz, moreno}, to which we refer the reader.


\section{Type II}
\label{sec:typeII}

Type II models are defined by $\det \lambda \neq 0$, from which follows $\det m = 0$.  The pseudomoduli space (\ref{flat}) is stable for
\begin{equation} \label{xminmaxII}
	|x| > x_{\text{min}},
\end{equation}
for some model dependent $x_{\text{min}}$.  Type II models have nonvanishing gaugino masses at leading order, and for this reason are phenomenologically promising \cite{eogm}.  At high temperatures, the only vacuum is inferred to be at origin of field space.  As the temperature drops, the supersymmetry and R-symmetry breaking EOGM vacuum (\ref{flat}) is cosmologically disfavored compared to the supersymmetric vacuum, unless the reheating temperature is significantly smaller than the messenger scale \cite{katz}.

The specific type II model we analyze is the same one considered at zero temperature in \cite{eogm}.  It is defined by the matrices
\begin{equation} \label{typeIImat}
	m_{ij} = \left(
		\begin{array}{cc}
			0 & m  \\
			0 & 0
		\end{array} \right),
	\qquad
	\lambda_{ij} = \left(
		\begin{array}{ccc}
			\lambda &  0 \\
			0 & \lambda
		\end{array} \right),
\end{equation}
which are seen to satisfy $\det \lambda \neq 0$ and $\det m = 0$.  Placing these into (\ref{eogmsp}) we obtain the superpotential
\begin{equation} \label{typeIIsp}
W = fx + m\phi_1 \widetilde{\phi}_2 + \lambda x (\phi_1 \widetilde{\phi}_1 + \phi_2 \widetilde{\phi}_2 ).
\end{equation}
From this superpotential we can see that this model is R-symmetric with R-charge assignments
\begin{equation}
\begin{gathered}
	R(x) = 2, \qquad
	R(\phi_1) = -R(\widetilde{\phi}_1) = \alpha, \\
	R(\phi_2) = - R(\widetilde{\phi}_2) = \alpha - 2,
\end{gathered}
\end{equation}
for arbitrary $\alpha$.  The treelevel scalar potential is
\begin{equation} \label{treeII}
\begin{split}
	V_0 &= |f + \lambda (\phi_1 \widetilde{\phi}_1 + \phi_2 \widetilde{\phi}_2)|^2 + |m \widetilde{\phi}_2 + \lambda x \widetilde{\phi}_1|^2 + |\lambda x {\phi}_1|^2 \\
		&\qquad +|\lambda x \widetilde{\phi}_2| ^2 +  |m \phi_1 + \lambda x \phi_2|^2.
\end{split}
\end{equation}
We assume all couplings to be real and positive without loss of generality, which can always be obtained by rotating the phases of the fields.

The scalar potential (\ref{treeII}) has supersymmetry breaking extrema at (\ref{flat}).  The $\phi_i = \widetilde{\phi}_i = 0$ direction is stable if the scalar mass squared matrix eigenvalues in this direction are all non-negative.  The general mass matrices for this model are given in section \ref{mBmFII} in the Appendix.  Unlike with the type I model, here the eigenvalues can be computed analytically.  Even so, they are complicated and unruly, but simplify dramatically in the small $y$ limit, where
\begin{equation} \label{ydefII}
	y \equiv \frac{\lambda f}{m^2}
\end{equation}
is a dimensionless quantity analogous to (\ref{rydefI}) for the type I model.  The $\phi_i = \widetilde{\phi}_i = 0$ direction is stable for
\begin{equation} \label{stablepspaceII}
	|x| > x_{\text{min}} = \frac{m}{\lambda}(2 y)^{1/3} + O(y).
\end{equation}
As expected, this stability condition is consistent with (\ref{xminmaxII}).

Just as with the type I model in the previous section, the analysis is simplified if we move to dimensionless quantities:
\begin{equation} \label{dimeqsII}
\begin{gathered}
	\overline{V} \equiv \frac{1}{m^4} V, \qquad
	\overline{x} \equiv \frac{\lambda}{m} x, \qquad
	\overline{\phi}_i \equiv \frac{\lambda}{m} \phi_i, \\
	\overline{\widetilde{\phi}}_i \equiv \frac{\lambda}{m} \widetilde{\phi}_i, \qquad
	\overline{T} \equiv \frac{1}{m} T,  \qquad
	\overline{\Lambda} \equiv \frac{1}{m} \Lambda, 	\\
	\overline{m}_S^2 \equiv \frac{1}{m^2} m_S^2, \qquad
	\overline{m}_F^2 \equiv \frac{1}{m^2} m_F^2,
\end{gathered}
\end{equation}
which are analogous to those for the type I model in (\ref{dimeqsI}).  In terms of these quantities the treelevel scalar potential (\ref{treeII}) simplifies to
\begin{equation} \label{treeII2}
\begin{split}
	\lambda^2 \overline{V}_0 &= |y +  \overline{\phi}_1 \overline{\widetilde{\phi}}_1 + \overline{\phi}_2 \overline{\widetilde{\phi}}_2|^2 + |\overline{\widetilde{\phi}}_2 + \overline{x} \overline{\widetilde{\phi}}_1|^2 + |\overline{x} \overline{\widetilde{\phi}}_2| ^2 \\
	&\qquad + | \overline{x} \overline{\phi}_1|^2 +  |\overline{\phi}_1 + \overline{x} \overline{\phi}_2|^2,
\end{split}
\end{equation}
and in the $\phi_i = \widetilde{\phi}_i = 0$ direction becomes $\overline{V}_0 = (y/\lambda)^2$.  The stability condition (\ref{stablepspaceII}) becomes $|\overline{x}| > (2y)^{1/3} + O(y)$ and the dimensionless mass matrices depend on $y$ and the dimensionless fields, but not $\lambda$.

The one-loop Coleman-Weinberg corrections (\ref{CWeqs2}) lift the flat direction (\ref{flat}) and we have a degenerate pseudomoduli space of supersymmetry breaking vacua in the $\phi_i = \widetilde{\phi}_i = 0$ direction parameterized by $x$, the pseudomodulus.  Unlike the type I model, however, the eigenvalues of the mass matrices for this model can be computed analytically and thus so too the one-loop scalar potential at zero temperature.  As already mentioned, the eigenvalues are complicated, but simplify in the small $y$ limit.  In fact, as we'll see in a moment, computable R-symmetry breaking vacua only exist for small $y$, so taking this limit is not constricting.  In this limit, $(\ref{CWeqsa2})$ gives \cite{eogm}
\begin{equation} \label{V1II}
\begin{split}
	\overline{V}^{(0)}_1 =& \frac{5 y^2}{16\pi^2} \Biggl[ \frac{1 + 12 |\bar{x}|^ 2}{1 + 4 |\bar{x}|^ 2} + \log \frac{|\bar{x}|^4}{\overline{\Lambda}^4} \\
	&+ \frac{1 + 2 |\bar{x}|^2}{(1 + 4 |\bar{x}|^2)^{3/2}} \log \frac{1 + 2|\bar{x}|^2 + \sqrt{ 1+ 4 |\bar{x}|^2}}{ 1 + 2 |\bar{x}|^2 - \sqrt{1+ 4 |\bar{x}|^2}} \Biggr].
\end{split}
\end{equation}
The zero temperature scalar potential, $\overline{V}^{(0)} = \overline{V}_0 + \overline{V}^{(0)}_1$, is plotted in figure \ref{fig:V1II}(f).  The integral for the finite temperature contribution (\ref{CWeqsb2}) will be done numerically.

In the $\phi_i = \widetilde{\phi}_i = 0$ direction, the treelevel contribution to the scalar potential, $\overline{V}_0 = (y/\lambda)^2$, is constant and does not affect the location of minima.  The zero temperature EOGM vacuum is therefore determined from (\ref{V1II}) alone, is independent of $y$ and $\lambda$ and is found to be at \cite{eogm}
\begin{equation} \label{min0II}
	\langle |\overline{x}|\rangle_0 = 0.249
\end{equation}
for $\overline{\Lambda} = 1$.  For this minimum to lie in the region (\ref{stablepspaceII}), where it can be reliably computed, we require roughly that $y < \langle |\bar{x}|\rangle_0^3 /2 = 0.00775$.  This is the promised result that the existence of computable R-symmetry breaking vacua requires $y$ to be small.  Below the exact numerical treatment shows that this bound is more accurately (see figure \ref{fig:pspaceII})
\begin{equation}
	y < 0.0058.
\end{equation}


\subsection{Numerical}

\begin{figure*}
		\includegraphics[width=6in]{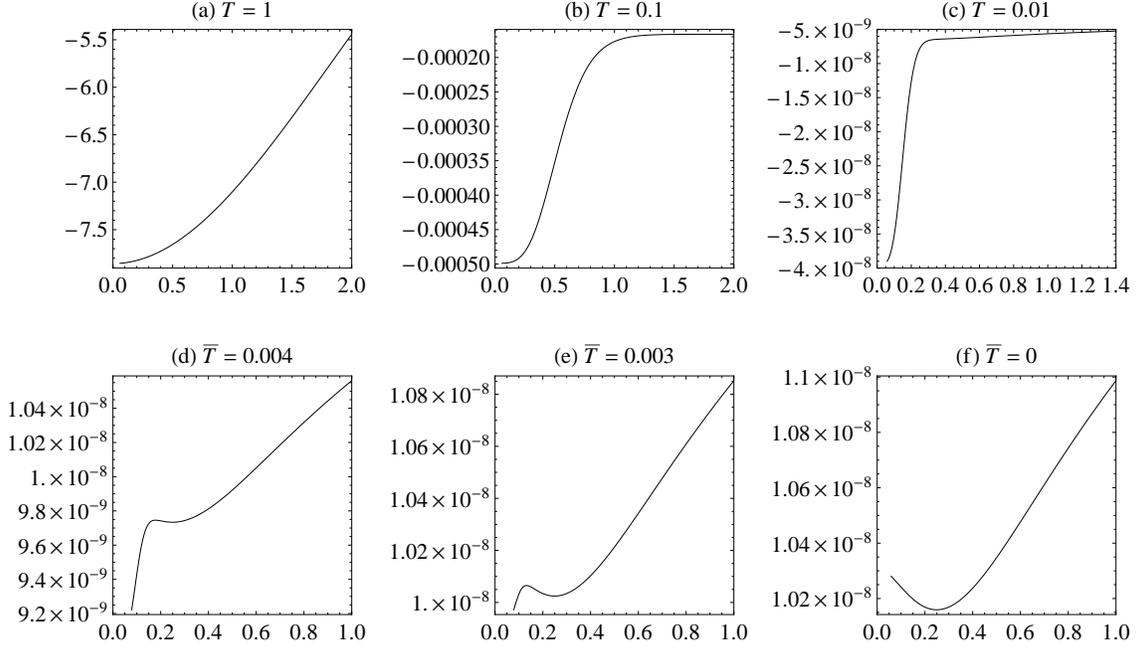}[h]
	\caption{$\overline{V}$ as a function of $|\bar{x}|$ is plotted for $y=10^{-4}$, $\lambda = 1$, cutoff $\overline{\Lambda}=1$ and various temperatures for the type II model.  The treelevel potential is stable and the one loop corrections are computable for $|\overline{x}| > \overline{x}_{\text{min}} = 0.0585$.  The zero temperature minimum of plot (f) is located at $\langle |\bar{x}| \rangle_0 = 0.249$.  The R-symmetry breaking minimum appears at critical temperature $\overline{T}_c = 0.0049$.  Neither of these quantities depend on $\lambda$. }
\label{fig:V1II}
\end{figure*}

For the $\phi_i = \widetilde{\phi}_i = 0$ direction, the treelevel contribution to the scalar potential, $\overline{V}_0 = (y/\lambda)^2$, is constant and does not affect the location of minima.  Still, as in the previous section, we include it for future comparison with the runaway direction and, for this reason, must specify a value for $\lambda$, which we again take to be $\lambda = 1$.  The location of minima in the $\phi_i = \widetilde{\phi}_i = 0$ depends on $\overline{x}$, $y$, $\overline{T}$ and the cutoff $\overline{\Lambda}$.  For a typical set of parameters ($y = 10^{-4}$, $\lambda = 1$ and $\overline{\Lambda} = 1$ \cite{eogm}) we have plotted $\overline{V}$ as a function of $|\bar{x}|$ for various temperatures in figure \ref{fig:V1II}.

Type II models are characterized by the $\phi_i = \widetilde{\phi}_i = 0$ direction being stable, i.e.~the eigenvalues of the scalar mass squared matrix are all nonnegative, for $|\overline{x}| > \overline{x}_{\text{min}}$.  Since these eigenvalues are used to calculate the Coleman-Weinberg correction, we can only reliably compute the one-loop scalar potential in this range.  As such, we do not have access to the origin of field space at one-loop, though we might be able to make plausible guesses.  Figure \ref{fig:V1II}(a) implies that the origin is the only minimum at very high temperatures, as is usually the case and which we will assume.  As the temperature is lowered, we see in figure \ref{fig:V1II}(d) a second minimum appears in addition to the one that still exists at the origin.  Further, this second minimum has a potential barrier between it and the origin.  Katz studied the existence of this  minimum analytically \cite{katz};  here we show an explicit plot.  As the temperature is lowered to zero, it is this second minimum that evolves to the R-symmetry breaking EOGM vacuum.  While it is not clear in figure \ref{fig:V1II}(f) whether the minimum at the origin persists down to zero temperature, Katz argues that it will \cite{katz}.  If it does, then if the system originates in the vacuum at the origin, the EOGM vacuum is disfavored.  The critical temperature for the appearance of the second minimum is
\begin{equation}
	\overline{T}_c = 0.0049. 
\end{equation}

\begin{figure}
	\includegraphics[width=2.5in]{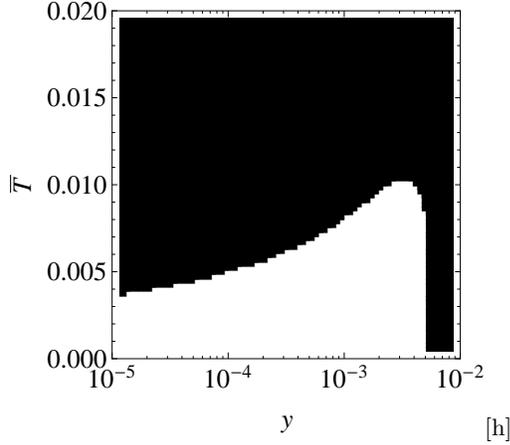}[h]
	\caption{$\overline{T}$ as function of $y$ is plotted for cutoff $\overline{\Lambda} =1$ for the type II model.  The white region is the computable parameter space in which an R-symmetry breaking minimum exists.}
\label{fig:pspaceII}
\end{figure}

The computable parameter space in which an R-symmetry breaking minimum exists in the $\phi_i = \widetilde{\phi}_i = 0$ direction is shown as the white region in figure \ref{fig:pspaceII}.
 We stress that figure \ref{fig:pspaceII} is the computable parameter space because we restricted our search to the region where the eigenvalues of the scalar mass squared matrix are all non-negative, approximately given by (\ref{stablepspaceII}), where the Coleman-Weinberg corrections at zero and finite temperature can be reliably computed.


\subsection{Runaway Direction}

In addition to the supersymmetry breaking vacuum in the $\phi_i = \widetilde{\phi}_i = 0$ direction, there is a supersymmetric vacuum at the end of a runaway direction given by
\begin{equation}
\begin{gathered}
	\overline{x} \rightarrow \frac{y}{2\overline{\phi}_2^{\,2}}, \qquad
	\overline{\phi}_1 \rightarrow - \frac{y}{2\overline{\phi}_2}, \qquad
	\overline{\phi}_2 \rightarrow \infty,\\
	\overline{\widetilde{\phi}}_1 \rightarrow \overline{\phi}_2, \qquad
	\overline{\widetilde{\phi}}_2 \rightarrow \overline{\phi}_1.
\end{gathered}
\end{equation}
To determine which vacuum is favored we employ the same method used with the type I model:  We find a single, stable path from vacua in the the $\phi_i = \widetilde{\phi}_i = 0$ direction, $\langle \overline{x} \rangle_T$, to the runaway direction.  We again use the method of \cite{moreno} for parameterizing this path and thus study
\begin{equation} \label{runpathII}
\begin{gathered}
	\overline{x} = h(\overline{\phi}_2)\frac{y}{2\overline{\phi}_2^{\,2}} + \left[1 - h(\overline{\phi}_2)\right] \langle \overline{x} \rangle_T, \quad
	\overline{\phi}_1 = -  h(\overline{\phi}_2)\frac{y}{2\overline{\phi}_2}, \\
	\overline{\widetilde{\phi}}_1 = \overline{\phi}_2, \qquad
	\overline{\widetilde{\phi}}_2 = \overline{\phi}_1,
\end{gathered}
\end{equation}
where $h(\overline{\phi}_2)$ is a function satisfying
\begin{equation}
	h(0) = 0, \qquad
	\lim_{\overline{\phi}_2 \rightarrow 0} \overline{\phi}_2^{\,-2} h(\phi_2) = 0, \qquad
	\lim_{\overline{\phi}_2 \rightarrow \infty} h(\overline{\phi}_2) = 1.
\end{equation}
For such a function, the path (\ref{runpathII}) begins at $\langle \overline{x} \rangle_T$ and ends at the runaway for $0 < \overline{\phi}_2 <\infty$.  There are many possible choices for $h(\overline{\phi}_2)$.  All choices we have tried leave some part of the path unstable \cite{arai}.  We choose
\begin{equation}
	h(\overline{\phi}_2) = \tanh\left(c\overline{\phi}_2^{\, d}\right),
\end{equation}
for some constant $c$ and $d > 2$.  Regardless of the values of $c$ and $d$ we find that the unstable region is always a segment of the path near $\overline{\phi}_2 = 0$.  Unfortunately, this is the most important region for determining the favored vacuum.  In the interest of having as much of this region be stable as possible, we use $c=10^3$ and $d=2.0001$.

\begin{figure*}
		\includegraphics[width=6in]{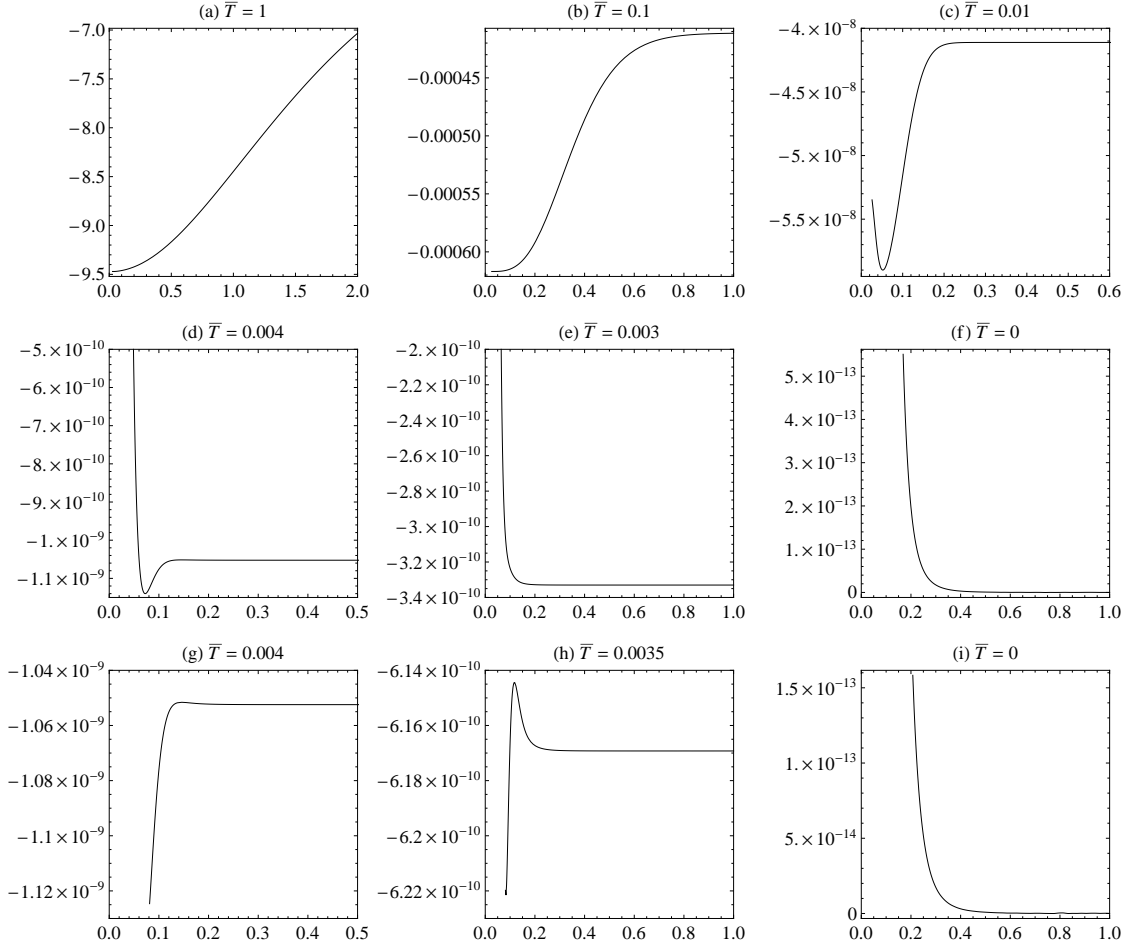}[h]
	\caption{The scalar potential, $\overline{V}$, along the path (\ref{runpathII}) is plotted as a function of $\overline{\phi}_2$ for $y=10^{-4}$, $\lambda = 1$, cutoff $\overline{\Lambda}=1$ and various temperatures for the type II model.  The top two rows use $\langle \bar{x} \rangle_T = 0$.  The bottom row uses $\langle \bar{x} \rangle_T \neq 0$ as determined from the bottom row of figure \ref{fig:V1II}.}
\label{fig:runawayII}
\end{figure*}

Having chosen a path, we can compute the loop corrected scalar potential along this path.  Using the same parameters ($y=10^{-4}$, $\lambda = 1$ and $\overline{\Lambda}=1$) as figure \ref{fig:V1II}, the result is shown in figure \ref{fig:runawayII}.
 In figure \ref{fig:runawayII} we have only plotted that part of the scalar potential that is stable and thus computable.  To understand these plots, compare them to figure \ref{fig:V1II}.  The top two rows of figure \ref{fig:runawayII} are made using $\langle \overline{x} \rangle_T = 0$.  The bottom row is made using $\langle \overline{x} \rangle_T \neq 0$, the exact value being the location of the R-symmetry breaking minima in the bottom row of figure \ref{fig:V1II}.

While we cannot reliably compute the scalar potential close to $\overline{\phi}_2=0$, we can infer from the top row of figure \ref{fig:runawayII} that the potential does not block the runaway direction.  Figure \ref{fig:runawayII}(c) shows the appearance of a minimum.  Once in this minimum, the system will escape to the runaway direction as seen in the second row.  On the other hand, if the system finds itself in the $\langle \overline{x} \rangle_T \neq 0$ vacuum in the $\phi_i = \widetilde{\phi}_i = 0$ direction, which is shown in the bottom row of figure \ref{fig:V1II}, then the bottom row of figure \ref{fig:runawayII} implies that the runaway direction is blocked.  In the previous subsection we computed the critical temperature for the $\langle \overline{x} \rangle_T \neq 0$ vacuum in the $\phi_i = \widetilde{\phi}_i = 0$ direction to appear and found it to be $\overline{T}_c = 0.0049$.  Figure \ref{fig:runawayII}(c) implies that for temperatures above $\overline{T}_c$ the runaway direction is favored.  We conclude, then, that the supersymmetry breaking EOGM vacuum is disfavored unless the reheating temperature is below $\overline{T}_c = 0.0049$.  This is in agreement with the general result in \cite{katz} that the reheating temperature be far below the messenger scale, which for this model is $m$.


\section{Type III}
\label{sec:typeIII}

Type III models are defined by $\det m = 0$ and $\det \lambda = 0$ and share features with both type I and type II models.  The pseudomoduli space (\ref{flat}) is stable for
\begin{equation} \label{xminmaxIII}
	x_{\text{min}} < |x| < x_{\text{max}},
\end{equation}
for model dependent $x_{\text{min}}$ and $x_{\text{max}}$.  Like type II models, type III models have nonvanishing gaugino masses at leading order and for this reason are phenomenologically promising \cite{eogm}.  At high temperatures, the only vacuum is inferred to be at the origin of field space.  If the Universe begins in this global minimum, the EOGM vacuum (\ref{flat}) is cosmologically disfavored, just as with type II models.  Unlike type II models, but similar to type I models, at high temperatures there might additionally be a local vacuum far from the origin.  If the Universe instead starts in this vacuum, then it is possible to end up in the EOGM vacuum  \cite{katz}.

The specific type III model we analyze is the same one considered at zero temperature in \cite{eogm}.  Katz studied it at finite temperature \cite{katz}, but not in the detail that we do here and not numerically.  It may be defined by the matrices
\begin{equation} \label{typeIIImat}
	m_{ij} = 
	\begin{pmatrix}
		m_1 & 0 & 0 & 0 \\
		0 & m_2 & 0 & 0 \\
		0 & 0 & m_1 & 0 \\
		0 & 0 & 0 & 0
	\end{pmatrix},
	\qquad
	\lambda_{ij} = 
	\begin{pmatrix}
	0 & \lambda & 0 & 0 \\
	0 & 0 & \lambda & 0 \\
		0 & 0 & 0 & 0\\
	0 & 0 & 0 & \lambda'
	\end{pmatrix},
\end{equation}
which are seen to satisfy $\det m = 0$ and $\det \lambda = 0$.  Placing these into (\ref{eogmsp}) we obtain the superpotential
\begin{equation}  \label{typeIIIsp}
\begin{split}
W &= fx + m_1 (\phi_1 \widetilde{\phi}_1 + \phi_3\widetilde{\phi}_3) + m_2 \phi_2 \widetilde{\phi}_2 \\
&\qquad+ \lambda x (\phi_1 \widetilde{\phi}_2 + \phi_2 \widetilde{\phi}_3) + \lambda' x \phi_4 \widetilde{\phi}_4,
\end{split}
\end{equation}
which is just the type I superpotential (\ref{typeIsp}) along with the $\lambda'$ term.  From this superpotential we can see that this model is R-symmetric with R-charge assignments
\begin{equation}
\begin{gathered}
	R(x) = 2, \qquad
	R(\phi_1) = - R(\widetilde{\phi}_2) = \alpha, \\
	R(\phi_2) = - R(\widetilde{\phi}_3) = 2 + \alpha,  \qquad
	R(\phi_3) = 4 + \alpha, \\
	R(\widetilde{\phi}_1) = 2 - \alpha, \qquad
	R(\phi_4) = -R(\widetilde{\phi}_4) = \beta,
\end{gathered}
\end{equation}
for arbitrary $\alpha$ and $\beta$.  The treelevel scalar potential is
\begin{equation} \label{treeIII}
\begin{split}
	V_0 &= |f + \lambda (\phi_1 \widetilde{\phi}_2 + \phi_2 \widetilde{\phi}_3) + \lambda' \phi_4 \widetilde{\phi}_4 |^2 + |m_1 \widetilde{\phi}_1 + \lambda x \widetilde{\phi}_2|^2 \\
	&\qquad + |m_1 \phi_1| ^2 
	+ |m_2 \widetilde{\phi}_2 + \lambda x \widetilde{\phi}_3|^2 + |m_2 \phi_2 + \lambda x \phi_1|^2 \\
	&\qquad+ |m_1 \widetilde{\phi}_3|^2 + |m_1 \phi_3 + \lambda x \phi_2|^2 \\
	&\qquad + |\lambda' x \phi_4|^2 + |\lambda' x \widetilde{\phi}_4|^2.
\end{split}
\end{equation}
We assume all couplings to be real and positive without loss of generality, which can always be obtained by rotating the phases of the fields.

The scalar potential (\ref{treeIII}) has supersymmetry breaking extrema at (\ref{flat}).  The $\phi_i = \widetilde{\phi}_i = 0$ direction is stable if the scalar mass squared matrix eigenvalues in this direction are all non-negative.  The general mass matrices for this model are given in section \ref{mBmFIII} in the Appendix.  The stability conditions are conveniently parameterized in terms of the dimensionless quantities
\begin{equation} \label{rsydefIII}
	r \equiv \frac{m_2}{m_1}, \qquad
	s \equiv \frac{\lambda'}{\lambda}, \qquad
	y \equiv \frac{\lambda f}{m_1 m_2}.
\end{equation} 
The $\phi_i = \widetilde{\phi}_i = 0$ direction is stable for
\begin{equation} \label{stablepspaceIII}
\begin{gathered}
	 y < 1, \qquad 
	 \frac{r}{s} < \frac{(1-y^2)^2}{4 y^3}, \\
	 \frac{m_1}{\lambda}\sqrt{\frac{y r}{s}} =x_{\text{min}} < |x| < x_{\text{max}} =  \frac{m_1}{\lambda}\frac{1-y^2}{2y}.
\end{gathered}
\end{equation}
As expected, these stability conditions are consistent with (\ref{xminmaxIII}).  The middle condition is just the statement $x_{\text{min}} < x_{\text{max}}$ and $x_{\text{max}}$ is the same as in the type I model (see (\ref{stablepspaceI})).  We note that these stability conditions are easily satisfied for small $y$.

As with the previous models, the analysis is simplified if we move to dimensionless quantities:
\begin{equation} \label{dimeqsIII}
\begin{gathered}
	\overline{V} \equiv \frac{1}{m_1^4} V, \qquad
	\overline{x} \equiv \frac{\lambda}{m_1} x, \qquad
	\overline{\phi}_i \equiv \frac{\lambda}{m_1} \phi_i, \\
	\overline{\widetilde{\phi}}_i \equiv \frac{\lambda}{m_1} \widetilde{\phi}_i, \qquad
	\overline{T} \equiv \frac{1}{m_1} T,  \qquad
	\overline{\Lambda} \equiv \frac{1}{m_1} \Lambda, 	\\
	\overline{m}_S^2 \equiv \frac{1}{m_1^2} m_S^2, \qquad
	\overline{m}_F^2 \equiv \frac{1}{m_1^2} m_F^2,
\end{gathered}
\end{equation}
which accompany those in (\ref{rsydefIII}).  The quantities (\ref{dimeqsIII}) are identical to those in (\ref{dimeqsI}) used with the type I model.  In terms of these quantities the treelevel scalar potential (\ref{treeIII}) simplifies to
\begin{equation} \label{treeIII2}
\begin{split}
	\lambda^2 V_0 &= |yr +  \overline{\phi_1} \overline{\widetilde{\phi}}_2 + \overline{\phi}_2 \overline{\widetilde{\phi}}_3 + s \overline{\phi}_4 \overline{\widetilde{\phi}}_4 |^2 + | \overline{\widetilde{\phi}}_1 +  \overline{x} \overline{\widetilde{\phi}}_2|^2 \\
	&\qquad + |\overline{\phi_1}| ^2 
	+ |r \overline{\widetilde{\phi}}_2 + \overline{x} \overline{\widetilde{\phi}}_3|^2 + |r \overline{\phi}_2 + \overline{x} \overline{\phi}_1|^2 + | \overline{\widetilde{\phi}}_3|^2 \\
	&\qquad + |\overline{\phi}_3 + \overline{x} \overline{\phi}_2|^2 + |s \overline{x} \overline{\phi}_4|^2 + |s \overline{x} \overline{\widetilde{\phi}}_4|^2,
\end{split}
\end{equation}
and in the $\phi_i = \widetilde{\phi}_i = 0$ direction becomes $\overline{V}_0 = (yr/\lambda)^2$.  The first two stability conditions in (\ref{stablepspaceIII}) stay the same while the third becomes $\sqrt{yr/s} < |\overline{x}| < (1-y^2)/2y$ and the dimensionless mass matrices depend on $r$, $s$, $y$ and the dimensionless fields, but not $\lambda$.

The one-loop Coleman-Weinberg corrections (\ref{CWeqs}) lift the flat direction (\ref{flat}) and we have a degenerate pseudomoduli space of supersymmetry breaking vacua in the $\phi_i = \widetilde{\phi}_i = 0$ direction parameterized by $x$, the pseudomodulus. The superpotential (\ref{typeIIIsp}) is the type I superpotential (\ref{typeIsp}) plus the $\lambda'$ term.  The one-loop Coleman-Weinberg correction separates into a piece dependent on $\lambda'$, or equivalently $s$, and a piece that does not, which is identical to the Coleman-Weinberg correction of the type I model \cite{eogm}:
\begin{equation} \label{V1separate}
	\overline{V}_1 = \overline{V}_{1,s} + \overline{V}_{1,\text{type I}}.
\end{equation}
Further, the $s$-dependent eigenvalues can be computed analytically (the $s$-independent eigenvalues cannot, as mentioned in our study of the type I model), yielding a zero temperature contribution
\begin{equation} \label{V1szero}
\begin{split}
	\overline{V}^{(0)}_{1,s} = \frac{5}{64\pi^2} s^2 \Biggl\{&2 (s|\bar{x}|^2+yr)^2\ln \left[ \frac{ s(s|x|^2 + yr)}{\overline{\Lambda}^2} \right] \\
	+&
	2 (s|\bar{x}|^2-yr)^2\ln \left[ \frac{ s( s|\bar{x}|^2 -yr)}{\overline{\Lambda}^2} \right] \\
	-&4s ^2 |\bar{x}|^4 \ln \left(\frac{s^2 |\bar{x}|^2}{\overline{\Lambda}^2} \right)
	\Biggr\}.
\end{split}
\end{equation}
Since the remaining eigenvalues cannot be computed analytically we have two options: An exact numerical analysis or introducing approximations such that an analytical analysis is possible.  As with the type I model, we pursue both options.


\subsection{Numerical}

\begin{figure*}
		\includegraphics[width=6in]{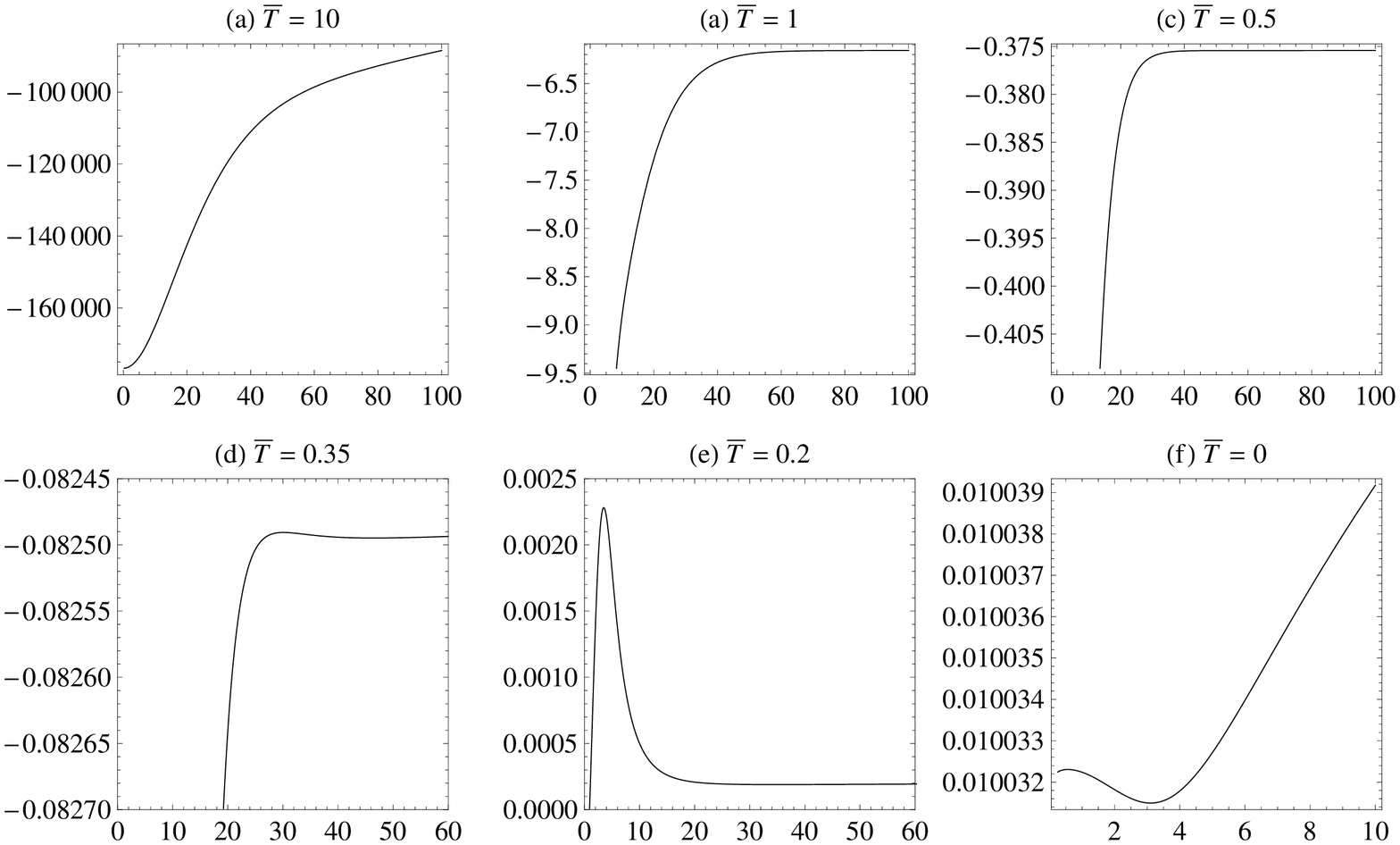}[h]
	\caption{$\overline{V}$ as a function of $|\bar{x}|$ is plotted for $r=10$, $s=0.15$, $y=10^{-3}$, $\lambda=1$, cutoff $\overline{\Lambda} = 1$ and various temperatures for the type III model.  The treelevel potential is stable and the one loop corrections are computable for $0.258 = \overline{x}_{\text{min}} < |\overline{x}| < \overline{x}_{\text{max}} = 500$.  The zero temperature minimum of plot (f) is located at $\langle |\bar{x}| \rangle_0 = 3.103$.  The R-symmetry breaking minimum appears at critical temperature $\overline{T}_c = 0.43$.  Neither of these quantities depend on $\lambda$.}
\label{fig:V1III}
\end{figure*}

For the $\phi_i = \widetilde{\phi}_i = 0$ direction we again retain the treelevel scalar potential, $\overline{V}_0 = (yr/\lambda)^2$, which in this direction is constant and does not affect the location of minima, for future comparison with the runaway direction.  For this reason we must specify a value for $\lambda$, which we again take to be $\lambda = 1$.  The location of minima in the $\phi_i = \widetilde{\phi}_i = 0$ direction depends on $\overline{x}$, $r$, $s$, $y$, $\overline{T}$ and the cutoff $\overline{\Lambda}$.  For a typical set of parameters ($r=10$, $s=0.15$, $y=10^{-3}$ and $\overline{\Lambda} = 1$ \cite{eogm}) we have plotted $\overline{V}$ as a function of $|\overline{x}|$ for various temperatures in figure \ref{fig:V1III}.

Type III models are characterized by the $\phi_i = \widetilde{\phi}_i = 0$ direction being stable, i.e.~the eigenvalues of the scalar mass squared matrix are all nonnegative, inside some range $\overline{x}_{\text{min}} < |\overline{x}| < \overline{x}_{\text{max}}$.  Since these eigenvalues are used to calculate the Coleman-Weinberg correction, we can only reliably compute the one-loop scalar potential in this range.  As such, we do not have access to the origin of field space at one-loop, though we might be able to make plausible guesses.  Figure \ref{fig:V1III}(a) implies that the origin is the only minimum at very high temperatures, as is usually the case and which we will assume.  As the temperature is lowered, we see in figure \ref{fig:V1III}(d) a second minimum appears in addition to the one that still exists at the origin.  Further, this second minimum has a potential barrier between it and the minimum at the origin.  As the temperature continues to lower it is this second minimum that evolves to the EOGM vacuum at zero temperature seen in figure \ref{fig:V1III}(f) \cite{katz}.  The R-symmetry breaking EOGM vacuum is located at
\begin{equation} \label{<x>0III}
	\langle |\bar{x}| \rangle_0 = 3.103.
\end{equation}
The minimum at the origin appears to persist down to zero temperature.  Since we can reliably determine a local maximum near the origin, there is a potential barrier between the vacuum at the origin and the EOGM vacuum even at zero temperature.  If the system originates in the global high temperature minimum at the origin, the EOGM vacuum is disfavored.  This is similar to the type II model.  The critical temperature for the appearance of the second minimum is
\begin{equation}
	\overline{T}_c = 0.43.
\end{equation}

\begin{figure*}
		\includegraphics[width=6in]{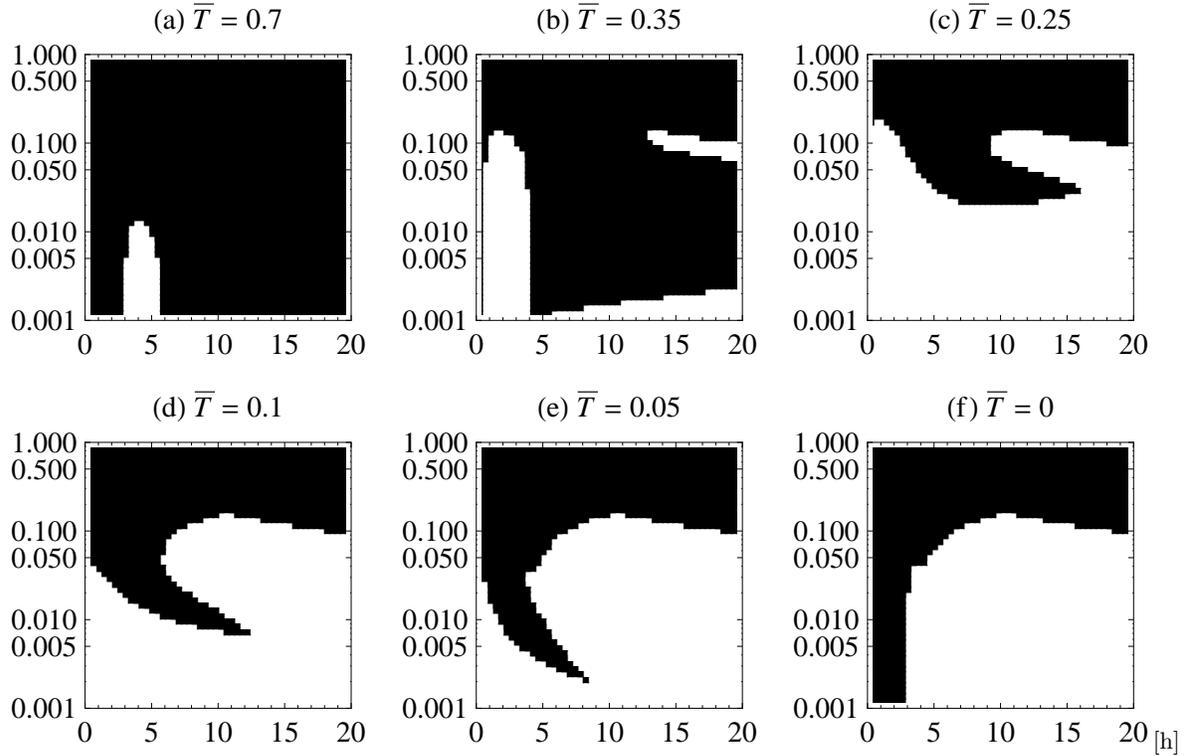}[h]
	\caption{$y$ as function of $r$ is plotted for $s=0.15$, cutoff $\overline{\Lambda} = 1$ and various temperatures for the type III model.  The white region is the computable parameter space in which an R-symmetry breaking minimum exists.}
\label{fig:pspaceIII}
\end{figure*}

The computable parameter space in which an R-symmetry breaking minimum exists in the $\phi_i = \widetilde{\phi}_i = 0$ direction is shown as the white region in figure \ref{fig:pspaceIII}.
 We stress that figure \ref{fig:pspaceIII} is the computable parameter space because we restricted our search to the region where the eigenvalues of the scalar mass squared matrix are all non-negative, given by (\ref{stablepspaceIII}), where the Coleman-Weinberg corrections at zero and finite temperature can be reliably computed.


\subsection{Analytical}

The one-loop contribution to the scalar potential (\ref{V1separate}) separates into a piece that depends on $\lambda'$, or equivalently $s$, and a piece that does not, which is identical to the type I one-loop contribution.  $V_{1,s}$ can be computed analytically at zero temperature and was given in (\ref{V1szero}).  The $s$-independent piece cannot because, as mentioned in our study of the type I model, the eigenvalues of the mass matrices cannot be computed analytically.  In the previous subsection we computed them numerically.  In this subsection we employ approximations and obtain approximate analytical results.

The approximations we employ are the same as used for the type I model and were worked out in section \ref{typeIanalytical}.  We consider the small supersymmetry breaking limit (by taking the small $y$ limit) and assume that we can use a truncated expansion of the effective K\"ahler potential to approximate $V^{(0)}_{1,\text{type I}}$.  The result was given in (\ref{rsymspI}):
\begin{equation}
	\overline{V}^{(0)}_{1,\text{type I}} \approx  -4(yr)^2 \bar{k}_{4}|\bar{x}|^2 -9(yr)^2 \bar{k}_{6}|\bar{x}|^4,
\end{equation}  
where $\bar{k}_4$ and $\bar{k}_6$ can be computed analytically without knowledge of the mass matrix eigenvalues and depend only on $r$.  They are given in (\ref{typeIk4k6}).  From (\ref{V1szero}), we have in the small $y$ limit
\begin{equation}
	\overline{V}^{(0)}_{1,s} = \frac{15}{32\pi^2} (yrs)^2 \left[ 1+ \frac{4}{3} \ln \left(\frac{s |\bar{x}|}{\overline\Lambda} \right) \right] + O(y^4).
\end{equation}
The zero temperature scalar potential will have an R-symmetry breaking vacuum, $\langle |\bar{x}|\rangle_0 \neq 0$, if
\begin{equation} \label{IIImin}
	\bar{k}_4 > 0, \qquad
	\bar{k}_6 < 0, \qquad
	s^2 < \frac{32\pi^2}{45} \frac{\bar{k}_4^2}{|\bar{k}_6|},
\end{equation}
and it will be located at
\begin{equation} \label{<x>III}
\begin{split}
	\langle |\bar{x}|^2 \rangle_0 &= \frac{1}{9} \frac{\bar{k}_4}{|\bar{k}_6|} \left( 1 + \sqrt{1 - \frac{45 s^2}{32\pi^2} \frac{|\bar{k}_6|}{\bar{k}^2_4}} \right) \\
	&= \frac{2}{9}\frac{\bar{k}_4}{|\bar{k}_6|} - \frac{5 s^2}{64\pi^2 \bar{k}_4} + O(s^4).
\end{split}
\end{equation}
As expected, the first term on the far right hand side of (\ref{<x>III}) is identical to $\langle |\bar{x}|^2 \rangle_0$ for the type I model in (\ref{keffminI2}), to which it reduces for $s=0$.  The inequality on the far right in (\ref{IIImin}) tells us that an R-symmetry breaking minimum will only occur for small $s$ (equivalently $\lambda' \ll \lambda$) \cite{eogm}.  For $r=10$ and $s=0.15$, (\ref{<x>III}) gives $\langle |\overline{x}|\rangle_0 = 3.36$, which agrees with the numerical result (\ref{<x>0III}) only in the first digit.  This is likely due to the truncated expansion becoming less reliable as $\langle |\overline{x}|\rangle_0$ gets larger.


\subsection{Supersymmetric Vacua}

In addition to the supersymmetry breaking vacuum in the $\phi_i = \widetilde{\phi}_i = 0$ direction, there is a supersymmetric vacuum at the end of a runaway direction.  It is easy to see that upon setting $\overline{\phi}_4 = \overline{\widetilde{\phi}}_4 = 0$, the type III scalar potential (\ref{treeIII2}) becomes the type I scalar potential (\ref{treeI2}).  A runaway direction is therefore given  by $\overline{\phi}_4 = \overline{\widetilde{\phi}}_4 \rightarrow 0$ and the type I runaway (\ref{rundirI}):
\begin{equation} \label{rundirIII}
\begin{gathered}
	\overline{x} \rightarrow \left( \frac{2}{y} \overline{\phi}_3^2 \right)^{1/3}, \qquad
	\overline{\phi}_1 \rightarrow \left(\frac{y^2 r^3}{4} \frac{1}{\overline{\phi}_3} \right)^{1/3}, \\
	\overline{\phi}_2 \rightarrow - \left(\frac{y}{2} \overline{\phi}_3 \right)^{1/3},  \qquad
	\overline{\phi}_3 \rightarrow \infty, \qquad
	\overline{\phi}_4 \rightarrow 0,\\
	\overline{\widetilde{\phi}}_1 \rightarrow \phi_3, \qquad
	\overline{\widetilde{\phi}}_2 \rightarrow \phi_2, \qquad
	\overline{\widetilde{\phi}}_3 \rightarrow \phi_1, \qquad
	\overline{\widetilde{\phi}}_4 \rightarrow \phi_4.
\end{gathered}
\end{equation}
The parameterization we use for a path connecting vacua in the $\phi_i = \widetilde{\phi}_i = 0$ direction to the runaway is just $\overline{\phi}_4 = \overline{\widetilde{\phi}}_4 = 0$ and the type I path (\ref{runpathI}):
\begin{equation} \label{runpathIII}
\begin{gathered}
	\overline{x} = \left( \frac{2}{y} \overline{\phi}_3^2 \right)^{1/3} + [1-h(\overline{\phi}_3)]\langle \overline{x} \rangle _T , \qquad
		\overline{\phi}_4 = 0\\
	\overline{\phi}_1  = h(\overline{\phi}_3) \left(\frac{y^2 r^3}{4} \frac{1}{\overline{\phi}_3} \right)^{1/3}, \quad
	\overline{\phi}_2 = - \left(\frac{y}{2} \overline{\phi}_3 \right)^{1/3},  \\
	\overline{\widetilde{\phi}}_1 = \overline{\phi}_3, \qquad
	\overline{\widetilde{\phi}}_2 = \overline{\phi}_2, \qquad
	\overline{\widetilde{\phi}}_3 = \overline{\phi}_1, \qquad
	\overline{\widetilde{\phi}}_4 = \overline{\phi}_4,
\end{gathered}
\end{equation}
where the function $h(\overline{\phi}_3)$ must satisfy (\ref{hreqI}) and we make the same choice (\ref{hI}) as made with the type I model:
\begin{equation}
	h(\overline{\phi}_3) = \tanh(c \overline{\phi}_3),
\end{equation}
using again $c=10$.  The path (\ref{runpathIII}) begins at $\langle |\overline{x}|\rangle_T$ and ends at the runaway for $0 < \phi_3 < \infty$.

\begin{figure*}
	\includegraphics[width=6in]{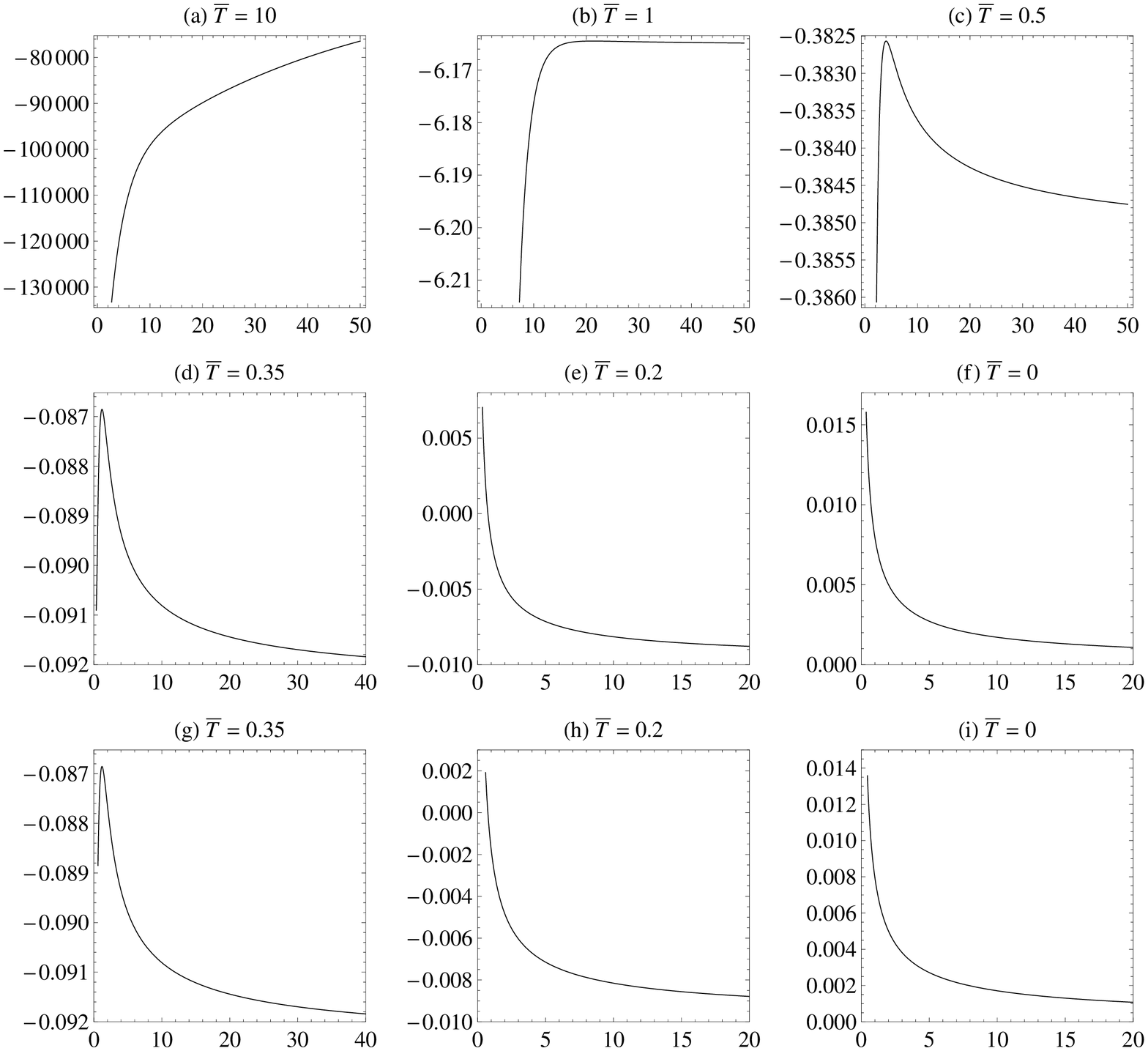}[h]
	\caption{The scalar potential, $\overline{V}$, along the path (\ref{runpathIII}) is plotted as a function of $\overline{\phi}_4$ for $r=10$, $s=0.15$, $y=10^{-3}$, $\lambda = 1$, cutoff $\overline{\Lambda}=1$ and various temperatures for the type III model.  The top two rows use $\langle \bar{x} \rangle_T = 0$.  The bottom row uses $\langle \bar{x} \rangle_T \neq 0$ as determined from the bottom row of figure \ref{fig:V1III}.}
\label{fig:runawayIII}
\end{figure*}

Having chosen a path, we can compute the loop corrected scalar potential along this path.  Using the same parameters ($r = 10$, $s=0.15$, $y=10^{-3}$, $\lambda = 1$ and $\overline{\Lambda} = 1$) as figure \ref{fig:V1III}, the result is shown in figure \ref{fig:runawayIII}.
  In figure \ref{fig:runawayIII} we have only plotted that part of the scalar potential that is stable and thus computable.  To understand these plots, compare them to figure \ref{fig:V1III}.  The top two rows of figure \ref{fig:runawayIII} are made using $\langle |\overline{x}| \rangle_T = 0$.  The bottom row is made using $\langle |\overline{x}| \rangle_T \neq 0$, the exact value being the location of the R-symmetry breaking minima in the bottom row of figure \ref{fig:V1III}.

While we cannot reliably compute the scalar potential close to $\overline{\phi}_3 = 0$, we might infer what is happening.  From figure \ref{fig:V1III} it appears that $\langle |\overline{x} | \rangle_T = 0$ will have lower potential energy than that part of the path to the runaway that is near it, which we are unable to plot in figure \ref{fig:runawayIII}.  This implies that the scalar potential must rise from $\langle |\overline{x} | \rangle_T = 0$ to that part plotted in figure \ref{fig:runawayIII}.  (This further implies that there should exist a local maximum in figures \ref{fig:runawayIII}(e), (f), (h) and (i), which we are unable to compute explicitly.)  The top and middle rows of figure \ref{fig:runawayIII} then imply that the runaway direction has a potential barrier when starting out at $\langle | \overline{x} | \rangle_T = 0$.  Similarly, if the system finds itself in the $\langle | \overline{x} | \rangle_T \neq 0$ vacuum in the $\phi_i = \widetilde{\phi}_i = 0$ direction, which is shown in the bottom row of figure \ref{fig:V1III}, the bottom row of figure \ref{fig:runawayIII} implies there is again a potential barrier blocking the runaway direction.  We conclude, then, that the runaway direction is disfavored.

\begin{figure}
	\includegraphics[width=3in]{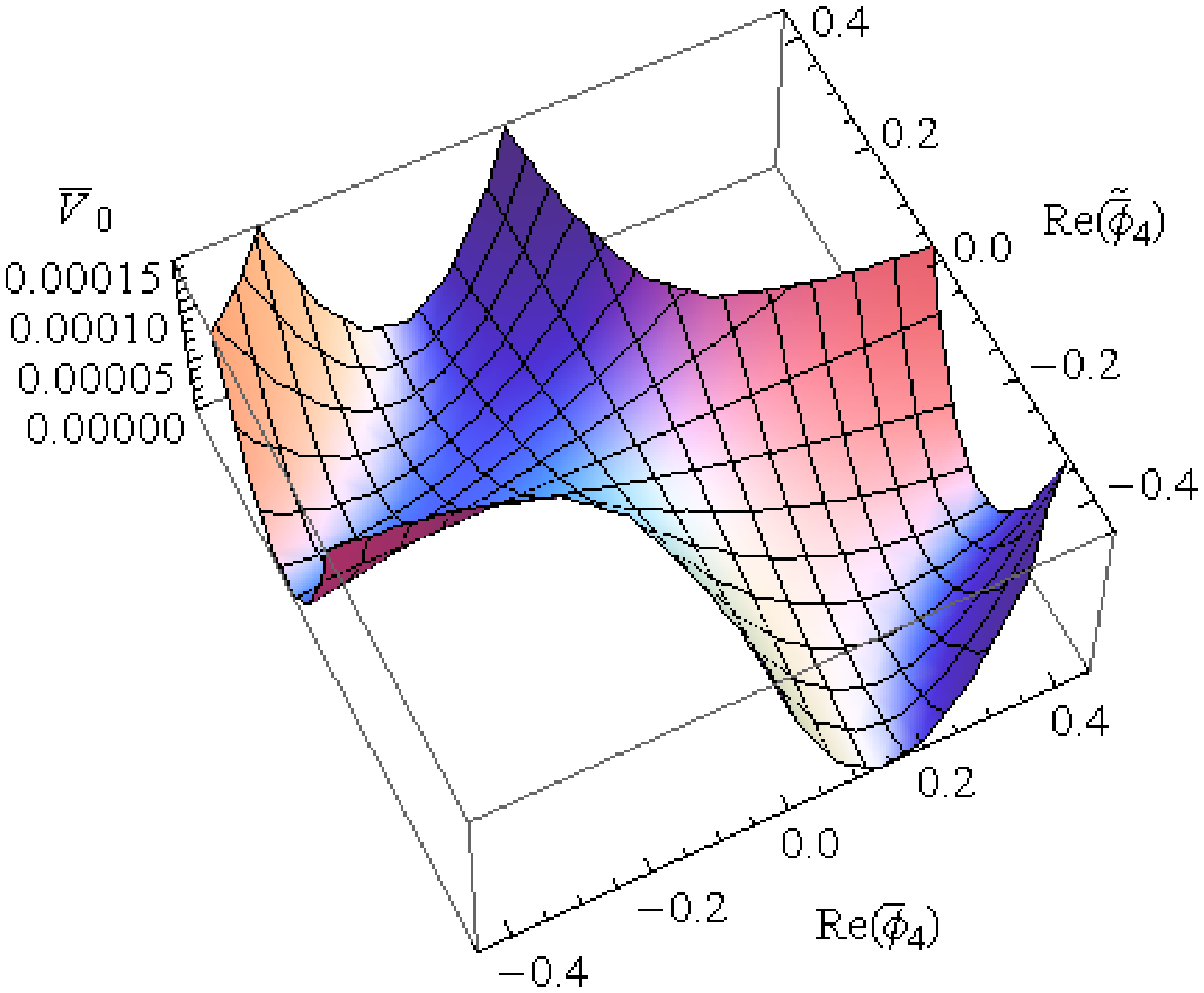}[h]
	\caption{The treelevel scalar potential, $\overline{V}_0$, is plotted with $\overline{x} = \overline{\phi}_{i\neq 4} = \overline{\widetilde{\phi}}_{i\neq 4} = 0$, $r=10$, $s=0.15$, $y=10^{-3}$ and $\lambda = 1$ for the type III model.  The supersymmetric vacua are located at $\overline{\phi}_4 \overline{\widetilde{\phi}}_4 = -yr/s = -0.0667$. }
\label{fig:susyvacIII1}
\end{figure}

In addition to a runaway direction there are supersymmetric vacua at
\begin{equation}
	\overline{x} = \overline{\phi}_{i\neq 4} = \overline{\widetilde{\phi}}_{i\neq 4} = 0, \qquad
	\overline{\phi}_4 \overline{\widetilde{\phi}}_4 = -\frac{yr}{s}.
\end{equation}
In trying to determine whether these vacua are favored or disfavored, we cannot employ the methods we have been using for the runaway directions.  The reason is that the scalar potential in the vicinity of the supersymmetric vacua is unstable (except, of course, at the vacua themselves, which are global minima) and we cannot compute the Coleman-Weinberg potential.  The (zero temperature) treelevel scalar potential can still be analyzed, however, and much can be learned from it.  If the system begins at the origin of field space then we can set $\overline{x} = \overline{\phi}_{i\neq 4} = \overline{\widetilde{\phi}}_{i\neq 4} = 0$ and study the treelevel scalar potential as a function of $\overline{\phi}_4$ and $\overline{\widetilde{\phi}}_4$.  This is shown in figure \ref{fig:susyvacIII1},
 from which we can see that the system will roll down to a supersymmetric vacuum.  If instead the system starts in the R-symmetry breaking vacuum, $\langle |x|\rangle_T \neq 0$, all fields but $x$ are at the origin of field space and we already know that the treelevel scalar potential is stable for this configuration and thus that there is a potential barrier blocking the system from rolling to the supersymmetric vacuum.  We conclude, then, that if the system is at the origin of field space, by zero temperature it will find itself in the supersymmetric vacuum, but if the system is in the R-symmetry breaking vacuum, $\langle |x|\rangle_T \neq 0$, it can find itself in the EOGM vacuum at zero temperature.  These conclusions agree with \cite{katz}.


\section{Conclusion}
\label{conclusion}

We studied R-symmetric minimal completions of EOGM at finite temperature.  These models, which are examples of direct gauge mediation, were classified into three types in \cite{eogm}.  For each type, after general remarks, we analyzed a specific model using both analytical and numerical techniques.

For the type I model, of which a related version had been studied previously at finite temperature \cite{moreno}, we found that in the $\phi_i = \widetilde{\phi}_i =0$ direction there is a single vacuum.  At high temperatures, it is at the origin of field space.  As the temperature is lowered toward zero, a second order phase transition occurs and it evolves to become the EOGM vacuum.  The type I model also had a runaway direction that contained a temperature dependent minimum.  If the system were to be in this minimum it would eventually escape to the runaway.  On the other hand, if the system were to begin at the origin of field space, we found that a potential barrier blocks the system from escaping to the runaway.  We concluded that the EOGM vacuum is the cosmologically preferred zero temperature vacuum.  While this is important for the viability of a model, type I models are considered somewhat pathological \cite{eogm}, since they have vanishing leading order gaugino masses.

For the type II model we found that in the $\phi_i = \widetilde{\phi}_i =0$ direction the origin of field space is probably the only minimum at very high temperatures and also probably persists down to zero temperature.  At some critical temperature a second minimum in this same direction, with $\langle |x|\rangle_T\neq 0$, appears.  If the system is in this second minimum, it will end up in the EOGM vacuum because there is a potential barrier blocking it from escaping to a runaway.  On the other hand, if the system is at the origin of field space, it will escape to the runaway.  Since the critical temperature for the appearance of the second minimum is lower than the temperature for the system to escape to the runaway from the origin of field space, we concluded that the system can only make it to the EOGM vacuum at zero temperature if the reheating temperature is lower than the critical temperature, and thus much lower than the messenger scale.

For the type III model we found that in the $\phi_i = \widetilde{\phi}_i =0$ direction the origin of field space is probably the only minimum at very high temperatures and also probably persists down to zero temperature.  At some critical temperature a second minimum in this same direction, with $\langle |x|\rangle_T\neq 0$, appears.  If the system is in this second minimum, it will end up in the EOGM vacuum because there is a potential barrier blocking it from escaping to a runaway.  So far, this is just like with the type II model.  If the system is instead at the origin of field space, there is again a potential barrier blocking the system from escaping to the runaway and it will eventually end up in a supersymmetric vacuum near the origin, unlike with the type II model.  The viability of this model, then, depends on the system being in the $\langle |x|\rangle_T\neq 0$ vacuum in the $\phi_i = \widetilde{\phi}_i =0$ and thus strongly depends on the initial conditions after inflation \cite{katz}.

To conclude, the type I model is able to end up in the EOGM vacuum the easiest, but has vanishing leading order gaugino masses.  The type II and III models do not have vanishing leading order gaugino masses, but the type II model has the most difficult time ending up in the EOGM vacuum, while the type III model can, but it depends strongly on the initial conditions after inflation.  Each model has its difficulties, but the type III model is the most promising of the three.


\section*{Acknowledgments}
A.~H.~was supported in part by a New Jersey Space Grant Consortium fellowship.  C.~M.~was supported in part by the Anthony and Renee Marlon Summer Research Fund.

\appendix

\section{Mass Matrices}

In this Appendix we give convenient formulas for computing the scalar and fermion mass squared matrices, which are used throughout this paper, and then use these formulas to compute the specific mass matrices for the models considered above.

The scalar and fermion mass squared matrices are given by
\begin{equation} \label{mBmFeqs}
\begin{split}
	m_B^2 &= \left(
		\begin{array}{cc}
			W_{ik}^\dag W^{kj} & W^\dag_{ijk}W^k \\
			W^{ijk}W_k^\dag & W^{ik}W_{kj}^\dag
		\end{array} \right), \\
	m_F^2 &= \left(
		\begin{array}{cc}
			W_{ik}^\dag W^{kj} & 0 \\
			0 & W^{ik}W_{kj}^\dag
		\end{array} \right),
\end{split}
\end{equation}
where $W$ is the superpotential.  Let $\phi$ be a column vector with components $\phi_i$ and $\widetilde{\phi}$ a column vector with components $\widetilde{\phi}_i$.  Defining the matrices
\begin{equation}
\begin{split}
	M &\equiv \left(
		\begin{array}{ccc}
			0 & 0 & 0 \\
			0 & m^T & 0 \\
			0 & 0 & m
		\end{array} \right)
	\qquad
	\Lambda \equiv 
		\left(
		\begin{array}{ccc}
			0 & 0 & 0 \\ 
			0 & \lambda^T & 0 \\
			0 & 0 & \lambda
		\end{array} \right)\\
	\qquad
	\Phi &\equiv \left(
		\begin{array}{ccc}
			0 & \widetilde{\phi}^T & \phi^T \\
			\phi & 0 & 0 \\
			\widetilde{\phi} & 0 & 0
		\end{array}\right)
	\\
	F &\equiv (f + \phi^T\lambda\tilde{\phi}) \left(
		\begin{array}{ccc}
			0 & 0 & 0 \\
			0 & 0 & 1 \\
			0 & 0 & 0
		\end{array} \right),
\end{split}
\end{equation}
then (\ref{mBmFeqs}) can be computed using
\begin{subequations} \label{mBmFformulas}
\begin{align} 
\begin{split}\label{ulformula}
	W^\dag_{ik}W^{kj} &= (W^{ik}W^\dag_{kj})^* \\
	&=  [(M + x\Lambda) + \Lambda\Phi + \Phi\Lambda]^\dag \\
	&\qquad \times[(M + x\Lambda) + \Lambda\Phi + \Phi\Lambda]
\end{split} \\
\begin{split} \label{urformula}
	W^\dag_{ijk}W^k &=
	(	W^{ijk}W_k^\dag)^* \\
	& = \left[F + \Phi^T(M + x\Lambda)^T\right] \Lambda^* \\
	&\qquad+ \left\{\left[F + \Phi^T(M + x\Lambda)^T\right] \Lambda^*\right\}^T.
\end{split}
\end{align}
\end{subequations}

For the specific models considered below, we write the matrices in terms of dimensionless quantities.  To reduce clutter, we do not include bars over the fields, as was done in the main body of the paper.  The matrices below are written with field order $(x,\, \phi_i, \, \widetilde{\phi}_i)$.


\subsection{Type I}
\label{mBmFI}

The specific type I model studied in section \ref{sec:typeI} has superpotential
\begin{equation}
	W = fx + m_1 (\phi_1 \widetilde{\phi}_1 + \phi_3\widetilde{\phi}_3) + m_2 \phi_2 \widetilde{\phi}_2 + \lambda x (\phi_1 \widetilde{\phi}_2 + \phi_2 \widetilde{\phi}_3).
\end{equation}
The matrices $m_{ij}$ and $\lambda_{ij}$ are written down in (\ref{typeImat}).  We assume all couplings to be real and positive without loss of generality.  In terms of the dimensionless quantities
\begin{equation}
\begin{gathered}
	r = \frac{m_2}{m_1}, \qquad 
	y = \frac{ \lambda f}{m_1 m_2}, \\
	x \rightarrow \frac{m_1}{\lambda}x, \qquad
	\phi_i \rightarrow \frac{m_1 }{\lambda}\phi_i, \qquad
	\widetilde{\phi}_i \rightarrow \frac{m_1 }{\lambda}\widetilde{\phi}_i,
\end{gathered}
\end{equation}
the formulas (\ref{mBmFformulas}) give
\begin{widetext}
\begin{equation}
	\frac{W_{ik}^\dag W^{kj}}{m_1^2}  = \left(
	\begin{array}{cccccccc}
	 D_{xx} & x \phi_1^* & r \phi_1^* +  x\phi_2^* &  \phi_2^* &  \widetilde{\phi}_2^* & r \widetilde{\phi}_3^* +  x \widetilde{\phi}_2^* &  x \widetilde{\phi}_3^*  \\
	 \phi_1 x^* & D_{11} & r x^* +  \widetilde{\phi}_2^* \widetilde{\phi}_3 & 0 & 0 &  \phi_1 \widetilde{\phi}_2^* &  \phi_2 \widetilde{\phi}_2^* \\
	r \phi_1 +  x^* \phi_2 & r x +  \widetilde{\phi}_2 \widetilde{\phi}_3^* & D_{22} &  x^* & 0 &  \phi_1 \widetilde{\phi}_3^* & \phi_2 \widetilde{\phi}_3^* \\
	 \phi_2 & 0 &  x & D_{33} & 0 &0 & 0 \\
	 \widetilde{\phi}_2 & 0 & 0 & 0  & D_{\tilde{1}\tilde{1}} & x & 0  \\
	r \widetilde{\phi}_3 +  x^* \widetilde{\phi}_2 &  \phi_1^* \widetilde{\phi}_2 &  \phi_1^* \widetilde{\phi}_3 & 0  &  x^* & D_{\tilde{2}\tilde{2}} & r x +  \phi_1^* \phi_2 \\
	 x^* \widetilde{\phi}_3 &  \phi_2^* \widetilde{\phi}_2 &  \phi_2^* \widetilde{\phi}_3 & 0 & 0 & r x^* +  \phi_1 \phi_2^* & D_{\tilde{3}\tilde{3}} 
	\end{array}	\right),
\end{equation}
with diagonal elements
\begin{subequations}
\begin{align}
	D_{xx} = \frac{1}{m_1^2} W_{xk}^\dag W^{kx} &= |\phi_1|^2 + |\phi_2|^2  + |\widetilde{\phi}_2|^2 + |\widetilde{\phi}_3|^2   \\
	D_{11} = \frac{1}{m_1^2} W_{1k}^\dag W^{k1} &= 1 + |x|^2 + |\widetilde{\phi}_2|^2 \\
	D_{22} = \frac{1}{m_1^2} W_{2k}^\dag W^{k2} &= r^2 + |x|^2 + |\widetilde{\phi}_3|^2 \\
	D_{33} = \frac{1}{m_1^2} W_{3k}^\dag W^{k3} &= 1 \\
	D_{\tilde{1}\tilde{1}} = \frac{1}{m_1^2} W_{\tilde{1}k}^\dag W^{k\tilde{1}} &= 1 \\
	D_{\tilde{2}\tilde{2}} = \frac{1}{m_1^2} W_{\tilde{2}k}^\dag W^{k\tilde{2}} &= r^2 + |x|^2 + |\phi_1|^2 \\
	D_{\tilde{3}\tilde{3}} = \frac{1}{m_1^2} W_{\tilde{3}k}^\dag W^{k\tilde{3}} &= 1 + |x|^2 + |\phi_2|^2,
\end{align}
\end{subequations}
and
\begin{equation}
	\frac{W^\dag_{ijk}W^k}{m_1^2}  = \left(
	\begin{array}{cccccccc}
		0 & r \phi_2 +  x \phi_1 & \phi_3 +  x \phi_2 & 0 & 0 &  \widetilde{\phi}_1 +  x \widetilde{\phi}_2 & r \widetilde{\phi}_2 +  x \widetilde{\phi}_3  \\
		r \phi_2 +  x \phi_1 & 0 & 0 & 0 &0 & R & 0 \\
		\phi_3 +  x \phi_2 & 0 & 0 & 0 & 0 & 0&  R \\
		0 & 0 & 0 & 0 & 0 & 0 & 0 \\
		0 & 0 & 0 & 0 & 0 & 0 & 0 \\
		 \widetilde{\phi}_1 +  x \widetilde{\phi}_2 & R & 0 & 0 & 0 & 0 & 0 \\
		r \widetilde{\phi}_2 +  x \widetilde{\phi}_3 & 0 & R & 0 & 0 & 0 & 0
	\end{array} \right),
\end{equation}
with  
\begin{equation}
	R = r y  + \phi_1 \widetilde{\phi}_2 + \phi_2 \widetilde{\phi}_3.
\end{equation}


\subsection{Type II}
\label{mBmFII}

The specific type II model studied in section \ref{sec:typeII} has superpotential
\begin{equation}
W = fx + m\phi_1 \widetilde{\phi}_2 + \lambda x (\phi_1 \widetilde{\phi}_1 + \phi_2 \widetilde{\phi}_2 ).
\end{equation}
The matrices $m_{ij}$ and $\lambda_{ij}$ are written down in (\ref{typeIImat}).  We assume all couplings to be real and positive without loss of generality.  In terms of the dimensionless quantities
\begin{equation}
	y = \frac{ \lambda f}{m^2}, \qquad
	x \rightarrow \frac{m}{\lambda}x, \qquad
	\phi_i \rightarrow \frac{m }{\lambda}\phi_i, \qquad
	\widetilde{\phi}_i \rightarrow \frac{m }{\lambda}\widetilde{\phi}_i,
\end{equation}
the formulas (\ref{mBmFformulas}) give
\begin{equation}
	\frac{W_{ik}^\dag W^{kj}}{m^2}  = \left(
	\begin{array}{ccccccc}
	 D_{xx}&  \phi_2^* + x\phi_1^* & x \phi_2^* &  x \widetilde{\phi}_1^* & \widetilde{\phi}_1^* +  x \widetilde{\phi}_2^* \\
	 \phi_2 +  x^* \phi_1  & D_{11} &  x +  \widetilde{\phi}_2 \widetilde{\phi}_1^* &  \phi_1 \widetilde{\phi}_1^* &  \phi_2 \widetilde{\phi}_1^* \\
	 x^* \phi_2 & x^* + \widetilde{\phi}_1 \widetilde{\phi}_2^* & D_{22} & \phi_1 \widetilde{\phi}_2^* & \phi_2 \widetilde{\phi}_2^* \\
	 x^* \widetilde{\phi}_1 &  \phi_1^* \widetilde{\phi}_1 &  \phi_1^* \widetilde{\phi}_2 &D_{\tilde{1}\tilde{1}} & x^* + \phi_1^* \phi_2 \\
	\widetilde{\phi}_1 + x^* \widetilde{\phi}_2 &   \phi_2^* \widetilde{\phi}_1 &  \phi_2^* \widetilde{\phi}_2 &  x +  \phi_1 \phi_2^* & D_{\tilde{2}\tilde{2}}
	\end{array}	\right),
\end{equation}
with diagonal elements
\begin{subequations}
\begin{align}
	D_{xx} = \frac{1}{m^2} W_{xk}^\dag W^{kx} &= |\phi_1|^2 + |\phi_2|^2 +  |\widetilde{\phi}_1|^2 + |\widetilde{\phi}_2|^2  \\
	D_{11} = \frac{1}{m^2} W_{1k}^\dag W^{k1} &= 1 + |x|^2 + |\widetilde{\phi}_1|^2 \\
	D_{22} = \frac{1}{m^2} W_{2k}^\dag W^{k2} &= |x|^2 + |\widetilde{\phi}_2|^2 \\
	D_{\tilde{1}\tilde{1}} = \frac{1}{m^2} W_{\tilde{1}k}^\dag W^{k\tilde{1}} &= |x|^2 + |\phi_1|^2 \\
	D_{\tilde{2}\tilde{2}} = \frac{1}{m^2} W_{\tilde{2}k}^\dag W^{k\tilde{2}} &= 1 + |x|^2 + |\phi_2|^2
\end{align}
\end{subequations}
and
\begin{equation}
	\frac{W^\dag_{ijk}W^k}{m^2}  = \left(
	\begin{array}{ccccc}
		0& x \phi_1 & \phi_1 + x \phi_2 & \widetilde{\phi}_2 +  x \widetilde{\phi}_1 &  x \widetilde{\phi}_2 \\
		x \phi_1 & 0 & 0 & R & 0 \\
		\phi_1 +  x \phi_2 & 0 & 0 & 0 &  R \\
		\widetilde{\phi}_2 +  x \widetilde{\phi}_1 & R & 0 & 0 & 0 \\
		 x \widetilde{\phi}_2 & 0 & R & 0 & 0
	\end{array} \right),
\end{equation}
with  
\begin{equation}
	R = y  + \phi_1 \widetilde{\phi}_1 + \phi_2 \widetilde{\phi}_2.
\end{equation}

\subsection{Type III}
\label{mBmFIII}

The specific type III model studied in section \ref{sec:typeIII} has superpotential
\begin{equation} 
W = fx + m_1 (\phi_1 \widetilde{\phi}_1 + \phi_3\widetilde{\phi}_3) + m_2 \phi_2 \widetilde{\phi}_2 + \lambda x (\phi_1 \widetilde{\phi}_2 + \phi_2 \widetilde{\phi}_3) \\
+ \lambda' x \phi_4 \widetilde{\phi}_4.
\end{equation}
The matrices $m_{ij}$ and $\lambda_{ij}$ are written down in (\ref{typeIIImat}).  We assume all couplings to be real and positive without loss of generality.  In terms of the dimensionless quantities
\begin{equation}
	r = \frac{m_2}{m_1}, \qquad
	s = \frac{\lambda'}{\lambda}, \qquad
	y = \frac{ \lambda f}{m_1 m_2}, \qquad
	x \rightarrow \frac{m_1}{\lambda}x, \qquad
	\phi_i \rightarrow \frac{m_1 }{\lambda}\phi_i, \qquad
	\widetilde{\phi}_i \rightarrow \frac{m_1 }{\lambda}\widetilde{\phi}_i,
\end{equation}
the formulas (\ref{mBmFformulas}) give
\begin{equation}
	\frac{W_{ik}^\dag W^{kj}}{m_1^2} = 
	\left(
	\begin{array}{ccccccccc}
	 D_{xx} & x \phi_1^* & r \phi_1^* +  x\phi_2^* &  \phi_2^* &s^2 x \phi_4^* &  \widetilde{\phi}_2^* & r \widetilde{\phi}_3^* +  x \widetilde{\phi}_2^* &  x \widetilde{\phi}_3^* & s^2 x \widetilde{\phi}_4^* \\
	 \phi_1 x^* & D_{11} & r x^* +  \widetilde{\phi}_2^* \widetilde{\phi}_3 & 0 &  s\widetilde{\phi}_4 \widetilde{\phi}_2^*  & 0 &  \phi_1 \widetilde{\phi}_2^* &  \phi_2 \widetilde{\phi}_2^* &   s\phi_4 \widetilde{\phi}_2^*\\
	r \phi_1 +  x^* \phi_2 & r x +  \widetilde{\phi}_2 \widetilde{\phi}_3^* & D_{22} &  x^* &  s\widetilde{\phi}_3^* \widetilde{\phi}_4 & 0 &  \phi_1 \widetilde{\phi}_3^* & \phi_2 \widetilde{\phi}_3^* &  s\phi_4 \widetilde{\phi}_3^* \\
	 \phi_2 & 0 &  x & D_{33} & 0 & 0 &0 & 0 & 0 \\
	s^2 x^* \phi_4 &  s\widetilde{\phi}_2 \widetilde{\phi}_4^* &  s \widetilde{\phi}_3 \widetilde{\phi}_4^* & 0 & D_{44} & 0 &  s\phi_1 \widetilde{\phi}_4^* &  s\phi_2 \widetilde{\phi}_4^* &  s^2\phi_4 \widetilde{\phi}_4^* \\
	 \widetilde{\phi}_2 & 0 & 0 & 0 & 0 & D_{\tilde{1}\tilde{1}} & x & 0 & 0 \\
	r \widetilde{\phi}_3 +  x^* \widetilde{\phi}_2 &  \phi_1^* \widetilde{\phi}_2 &  \phi_1^* \widetilde{\phi}_3 & 0 &  s\phi_1^* \widetilde{\phi}_4 &  x^* & D_{\tilde{2}\tilde{2}} & r x +  \phi_1^* \phi_2 & s\phi_1^* \phi_4 \\
	 x^* \widetilde{\phi}_3 &  \phi_2^* \widetilde{\phi}_2 &  \phi_2^* \widetilde{\phi}_3 & 0 &  s\phi_2^* \widetilde{\phi}_4 & 0 & r x^* +  \phi_1 \phi_2^* & D_{\tilde{3}\tilde{3}} &  s\phi_2^* \phi_4 \\
	s^2 x^* \widetilde{\phi}_4 &  s\phi_4^* \widetilde{\phi}_2 &  s\phi_4^* \widetilde{\phi}_3 & 0 & s^2\phi_4^* \widetilde{\phi}_4 & 0 &  s\phi_1 \phi_4^* &  s\phi_2 \phi_4^* & D_{\tilde{4}\tilde{4}}
	\end{array}	\right),
\end{equation}
with diagonal elements
\begin{subequations}
\begin{align}
	D_{xx} = \frac{1}{m_1^2} W_{xk}^\dag W^{kx} &= |\phi_1|^2 + |\phi_2|^2 + |\phi_4|^2 + |\widetilde{\phi}_2|^2 + |\widetilde{\phi}_3|^2   + |s\widetilde{\phi}_4|^2\\
	D_{11} = \frac{1}{m_1^2} W_{1k}^\dag W^{k1} &= 1 + |x|^2 + |\widetilde{\phi}_2|^2 \\
	D_{22} = \frac{1}{m_1^2} W_{2k}^\dag W^{k2} &= r^2 + |x|^2 + |\widetilde{\phi}_3|^2 \\
	D_{33} = \frac{1}{m_1^2} W_{3k}^\dag W^{k3} &= 1 \\
	D_{44} = \frac{1}{m_1^2} W_{4k}^\dag W^{k4} &= s^2(|x|^2 + |\widetilde{\phi}_4|^2) \\
	D_{\tilde{1}\tilde{1}} = \frac{1}{m_1^2} W_{\tilde{1}k}^\dag W^{k\tilde{1}} &= 1 \\
	D_{\tilde{2}\tilde{2}} = \frac{1}{m_1^2} W_{\tilde{2}k}^\dag W^{k\tilde{2}} &= r^2 + |x|^2 + |\phi_1|^2 \\
	D_{\tilde{3}\tilde{3}} = \frac{1}{m_1^2} W_{\tilde{3}k}^\dag W^{k\tilde{3}} &= 1 + |x|^2 + |\phi_2|^2 \\
	D_{\tilde{4}\tilde{4}} = \frac{1}{m_1^2} W_{\tilde{4}k}^\dag W^{k\tilde{4}} &= s^2 (|x|^2 + |\phi_4|^2)
\end{align}
\end{subequations}
and
\begin{equation}
	\frac{W^\dag_{ijk}W^k}{m_1^2}  = 
	\left(
	\begin{array}{ccccccccc}
		0 & r \phi_2 +  x \phi_1 & \phi_3 +  x \phi_2 & 0 & s^2 x \phi_4 & 0 &  \widetilde{\phi}_1 +  x \widetilde{\phi}_2 & r \widetilde{\phi}_2 +  x \widetilde{\phi}_3 & s^2 x \widetilde{\phi}_4 \\
		r \phi_2 +  x \phi_1 & 0 & 0 & 0 & 0 &0 & R & 0 & 0 \\
		\phi_3 +  x \phi_2 & 0 & 0 & 0 & 0 & 0 & 0&  R & 0 \\
		0 & 0 & 0 & 0 & 0 & 0 & 0 & 0 & 0\\
		 s^2 x \phi_4 & 0 & 0 & 0 & 0 & 0 & 0 & 0 & sR \\		
		0 & 0 & 0 & 0 & 0 & 0 & 0 & 0 & 0 \\
		 \widetilde{\phi}_1 +  x \widetilde{\phi}_2 & R & 0 & 0 & 0 & 0 & 0 & 0 & 0\\
		r \widetilde{\phi}_2 +  x \widetilde{\phi}_3 & 0 & R & 0 & 0 & 0 & 0 & 0 & 0\\
		s^2 x \widetilde{\phi}_4 & 0 & 0 & 0 & s R & 0 & 0 & 0 & 0
	\end{array} \right),
\end{equation}
\end{widetext}
with
\begin{equation}
	R =  r y  + \phi_1 \widetilde{\phi}_2 + \phi_2 \widetilde{\phi}_3 + s \phi_4 \widetilde{\phi}_4.
\end{equation}

\end{document}